\documentclass[aps,prx,twocolumn,groupedaddress]{revtex4-1}

\usepackage{amsmath}
\usepackage{graphicx}
\usepackage{dsfont}
\usepackage{xcolor}
\usepackage{amsfonts}
\usepackage{ulem}
\usepackage{enumitem}  
\usepackage{bbm}
\newcommand{\Bra}[1]{\left\langle #1 \right|}
\newcommand{\Ket}[1]{\left| #1\right\rangle}
\newcommand{\avg}[1]{\langle #1\rangle}

\definecolor{colorlibk}{rgb}{0.05, 0.4, 0.3} 

\DeclareMathOperator{\Tr}{Tr}
\begin{document}

\title{Discrete Time Crystal in the Gradient Field Heisenberg Model}

\author{Bikun Li, John S. Van Dyke, Ada Warren, Sophia E. Economou, Edwin Barnes}

\affiliation{Department of Physics, Virginia Tech, Blacksburg, Virginia 24061, USA}

\begin{abstract}
		We show that time crystal phases, which are known to exist for disorder-based many-body localized systems, also appear in systems where localization is due to strong magnetic field gradients. Specifically, we study a finite Heisenberg spin chain in the presence of a gradient field, which can be realized experimentally in quantum dot systems using micromagnets or nuclear spin polarization. Our numerical simulations reveal time crystalline order over a broad range of realistic quantum dot parameters, as evidenced by the long-time preservation of spin expectation values and the asymptotic form of the mutual information. We also consider the undriven system and present several diagnostics for many-body localization that are complementary to those recently studied. Our results show that these non-ergodic phases should be realizable in modest-sized quantum dot spin arrays using only demonstrated experimental capabilities.
\end{abstract}
	
	\maketitle
	
	\section{Introduction}
	
	Nonequilibrium phases of matter have recently drawn much attention within the field of condensed matter physics, owing in part to substantial progress in the precise experimental control of quantum systems. Cold atom and trapped ion setups have enabled the realization of numerous many-body Hamiltonians, with a high degree of tunability for interaction strengths and other model parameters \cite{Bloch2008,Lewenstein2012,Kim2010}. Protocols for sudden quenches or periodic driving can generically produce far-from-equilibrium states of these systems, which are not easily accessible in traditional solid-state contexts. This allows the detailed study of previously unexplored regimes of important models, and may be of use in the development of quantum information technologies \cite{Sorensen2001}.
	
	Gate-defined quantum dots (QDs) provide another platform for the simulation of interesting Hamiltonians. In these systems, individual electrons are confined to a semiconductor interface and laterally trapped by adjusting gate voltages. Here too, the rapid development of experimental control over multi-dot arrays holds promise for ultrafast initialization, evolution, and readout for quantum simulations, with good isolation from the environment and tunability of parameters \cite{Nichol2017,Takumi2018,Kandel2019,Mills2019,Sigillito2019,Dehollain2019,Yang2019,Huang2019,Xue2019,Cerfontaine2019}. Unlike other systems that have been studied to date, QD spin arrays naturally realize Heisenberg spin chains due to nearest-neighbor exchange couplings. Thus, these systems offer the opportunity to study new types of nonequilibrium many-body physics. 
	
	On the theoretical side, limitations of the framework of equilibrium statistical mechanics have been identified. In particular, the phenomenon of many-body localization (MBL) shows that a quantum system may fail to provide a bath for its own subsystems, thereby evading thermalization \cite{Basko2006,Pal2010,Nandkishore2015,Abanin2019}. Periodically-driven MBL systems have been shown to host exotic nonequilibrium phases of matter, both symmetry-breaking and topological. In particular, it was recently proposed \cite{Else2016,Yao2017} and experimentally observed \cite{Zhang2017,Choi2017} that driven MBL systems can spontaneously break the discrete time translation symmetry of the drive, which was argued to be impossible in equilibrium \cite{Watanabe2015}. The breaking of this symmetry led to the observation of discrete time crystal (DTC) phases for the first time, as signified by observables exhibiting oscillatory behavior at a frequency that differs from the drive.
	
	In most examples of non-ergodic phases discovered to date, strong on-site disorder plays a key role. Recently, magnetic field gradients in 1D spin models (or equivalently, via Jordan-Wigner transformation, spinless fermions in an electric field) have been found to exhibit similar signatures as disorder-based MBL systems \cite{Schulz2019,vanNieuwenburg2019,Wu2019,Taylor2019}. These features persist for weak or even no disorder in the magnetic field. This naturally raises the question of whether a gradient field model can host a DTC phase when periodically driven.
	
	In this paper, we demonstrate the existence of the gradient-field Heisenberg time crystal through a combination of analytical arguments and numerical simulations. We find that signatures of this phase are evident in systems containing as few as four sites and in parameter regimes consistent with current QD experiments. We also present alternative diagnostics of MBL in the absence of driving to further clarify the relationship between disorder-induced and gradient-induced localization in 1D spin chains.
	
	The paper is organized as follows. In Section \ref{sec:model} we introduce the gradient-field Heisenberg model and the periodic driving which leads to the stabilization of the DTC. In Section \ref{sec:DTC} we present evidence for the time crystal phase in this system, which includes the long-time preservation of spin expectation values and the persistence of mutual information between distant spins in the Floquet eigenstates. In Section \ref{sec:MBL} we discuss signatures of MBL in the undriven model, including the quantum Fisher information and the absorption of energy under weak driving. Many of the numerical simulations in Sections \ref{sec:DTC} and \ref{sec:MBL} were performed using the QuSpin exact diagonalization Python library \cite{Weinberg2017}. We summarize our conclusions in Section \ref{sec:conclusions}.

	\section{Model \label{sec:model}}
\begin{figure}
		\includegraphics[scale=0.15]{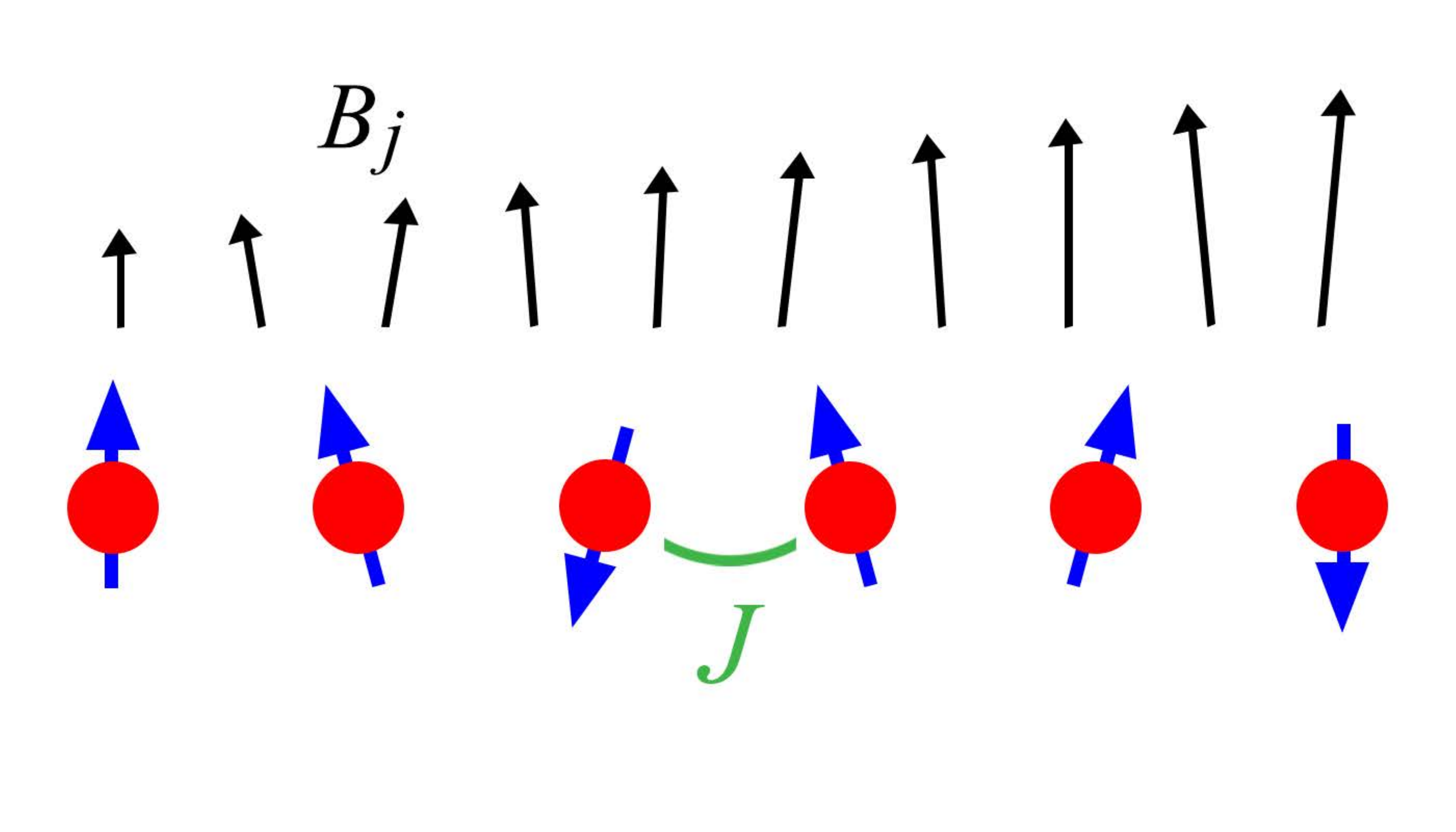}
		\caption{Schematic illustration of the model of Eq.~\eqref{eq:heisenberg}.  Electron spins reside on a linear chain of QDs, and interact via a Heisenberg coupling $J$.  They are also subject to a magnetic field gradient across the chain, produced by an embedded micromagnet.}
		\label{fig:cartoon}
	\end{figure}

	We consider a nearest-neighbor spin-1/2 Heisenberg model in 1D with an onsite magnetic field that includes an applied field, a random disorder term, and a gradient term:
	\begin{equation}\label{eq:heisenberg}
	\begin{aligned}
	\mathcal{H}_{H} &= \frac{J}{4} \sum_{j=1}^{L-1} \vec{\sigma}_j \cdot \vec{\sigma}_{j+1} + \frac{1}{2}\sum_{j=1}^{L} B_j  \sigma^z_j\;,\\
	\vec{B}_j &= B_0 + g(j-1) + \delta B_j,
	\end{aligned}
	\end{equation}
	where $\vec{\sigma}_j=(\sigma^x_j,\sigma^y_j,\sigma^z_j)$ is a vector of Pauli matrices. The system is illustrated in Fig.~\ref{fig:cartoon}. Here $L$ is the number of sites in the chain, $J$ is the exchange coupling between electron spins, $B_0$ is the applied magnetic field, $g$ is the strength of the field gradient, and $\delta B_j$ is a random value drawn from a Gaussian distribution with zero mean and standard deviation $\sigma_B$ (we typically consider $\sigma_B\ll g$). Experimentally, the field disorder in GaAs QDs arises from the nuclear spin bath, and occurs along all directions (not just along $\hat{z}$). However, in the presence of a large external applied field, the total field (applied plus noise) remains approximately parallel to $\hat{z}$, and the primary effect of the nuclear bath is to change the magnitude of the field at each dot \cite{Neder2011,Malinowski2017}. 
	
	To study the DTC phase of this model, we introduce a driving term which approximately flips all the spins after each period $T$. In the idealized case where the pulse is approximated as a delta function, the driving term is
	\begin{equation}\label{eq:deltapulse}
	V_\delta(t) = \left(\frac{\pi}{2} - \epsilon\right)  \sum_{s=1}^\infty \delta (t - s T ) \sum_{j=1}^L \sigma^x_j,
	\end{equation}
	where $\epsilon$ is the pulse error. While the delta-pulse driving generates the essential features of the gradient field DTC, finite-amplitude pulses are of course used in experiments. To determine the impact of this on the DTC, we will also consider the more realistic case of electric dipole spin resonance (EDSR) pulses, which are employed to execute single-spin rotations in QD systems \cite{Nowack2007,Laird2007,Pioro-Ladriere2008}:
	\begin{equation}
	\begin{aligned}
	V_{\mathrm{EDSR}}(t) &= 2(\eta T)^{-1}\left(\frac{\pi}{2} - \epsilon\right) \sum_{j=1}^L\sigma_j^x \cos(\Omega_j t)\;,\\
	&\quad
	\mathrm{for}\quad
	sT-\eta T < t < sT,\;s\in\mathbb{Z}^+\;,\\
	V_{\mathrm{EDSR}}(t) &= 0,\;\;\mathrm{otherwise},
	\end{aligned}
	\end{equation} 
	where $\Omega_j = B_0 + g(j-1)$ is the control frequency, and the duty cycle $\eta$ controls the pulse duration, with $ (\Omega_j T)^{-1}\ll\eta < 1$. In what follows, we find that the delta-pulse and EDSR-pulse models give qualitatively similar results. Thus, we will primarily focus on the delta-pulse model, and we will specify the occasions on which we switch to EDSR pulses to check the quantitative differences that occur in that case.
	
	\section{Discrete Time Crystal \label{sec:DTC}}
	The concept of a time crystal, that is, a state which spontaneously breaks time translation symmetry, has been hotly debated \cite{Bruno2013,Wilczek2013,Watanabe2015,Valerii2019} following its initial proposal \cite{Wilczek2012}. The difficulties arising in its formulation were sidestepped by considering periodically driven systems, for which a precise definition of a DTC was put forward in Ref. \cite{Else2016}. For a periodic Hamiltonian $\mathcal{H}(t+T) = \mathcal{H}(t)$, a DTC occurs when every short-range correlated state (i.e. one for which cluster decomposition holds) breaks the time translation symmetry of the drive for the expectation value of some observable, $\langle \mathcal{O}(t+T) \rangle \neq \langle \mathcal{O}(t) \rangle$. 
	
	In Ref. \cite{Barnes2019}, time crystalline behavior was studied in finite QD spin chains, with an emphasis on the state preservation properties of the phase. In particular, for the 1D Ising model with field disorder, the expectation value $\langle \sigma^z_1 (t) \rangle$ at the end of the chain was found to be well-preserved when viewed stroboscopically at times $t = 2sT$, where $s$ is an integer. This signifies the existence of a robust time crystal phase that persists over a broad range of interaction strengths and pulse errors.
	
	On the other hand, for the Heisenberg model no time crystal phase was observed within the same range of parameters (except for large disorder $\sigma_B/J > 10^3$, which is hard to achieve experimentally).  However, applying a special ``Heisenberg to Ising" (``H2I'') pulse  sequence to the spin chain was found to counteract the transverse terms in the Heisenberg coupling, thereby converting the model to an Ising-like form and restoring the DTC phase \cite{Barnes2019}. The effectiveness of this approach scales with the number of H2I pulses applied during a Floquet period. This becomes experimentally challenging; for instance in QD systems, as many as $\sim 100$ H2I pulses per period (in addition to the usual driving pulses) would be needed to completely restore the DTC. Thus, there is strong motivation for finding alternative methods of stabilizing the DTC in the Heisenberg model.  
	
	Applying a magnetic field gradient to the QD chain provides such an alternative. A gradient field profile can be obtained experimentally by embedding a micromagnet of the appropriate geometry beneath the chain.  In fact, such magnets are already used to provide single-spin addressability, as required for quantum information processing \cite{Watson2018,Sigillito2019,Takeda2019}. Alternatively, in the case of GaAs QDs, nuclear spin programming can be used to create large gradient fields \cite{Foletti2009,Bluhm2010,Nichol2017}. A suitable diagnostic of the DTC phase of the model $\mathcal{H}_{H} + V_\delta$ is the expectation value of the $z$ component of the first spin, averaged over both time and disorder realizations: $\langle \langle \sigma^z_j(2sT) \rangle \rangle = \frac{1}{s_{\max}} \sum_{s=0}^{s_{\max}-1} \Bra{\psi_i} \sigma^z_j(2sT)   \Ket{\psi_i}$, where $\Ket{\psi_i}$ is the initial state of the spin chain. Here, we only include the values at every second Floquet period, and we take $s_{\max}=200$, corresponding to 400 periods of the driving field. 
	
\begin{figure}
		\includegraphics[scale=0.65]{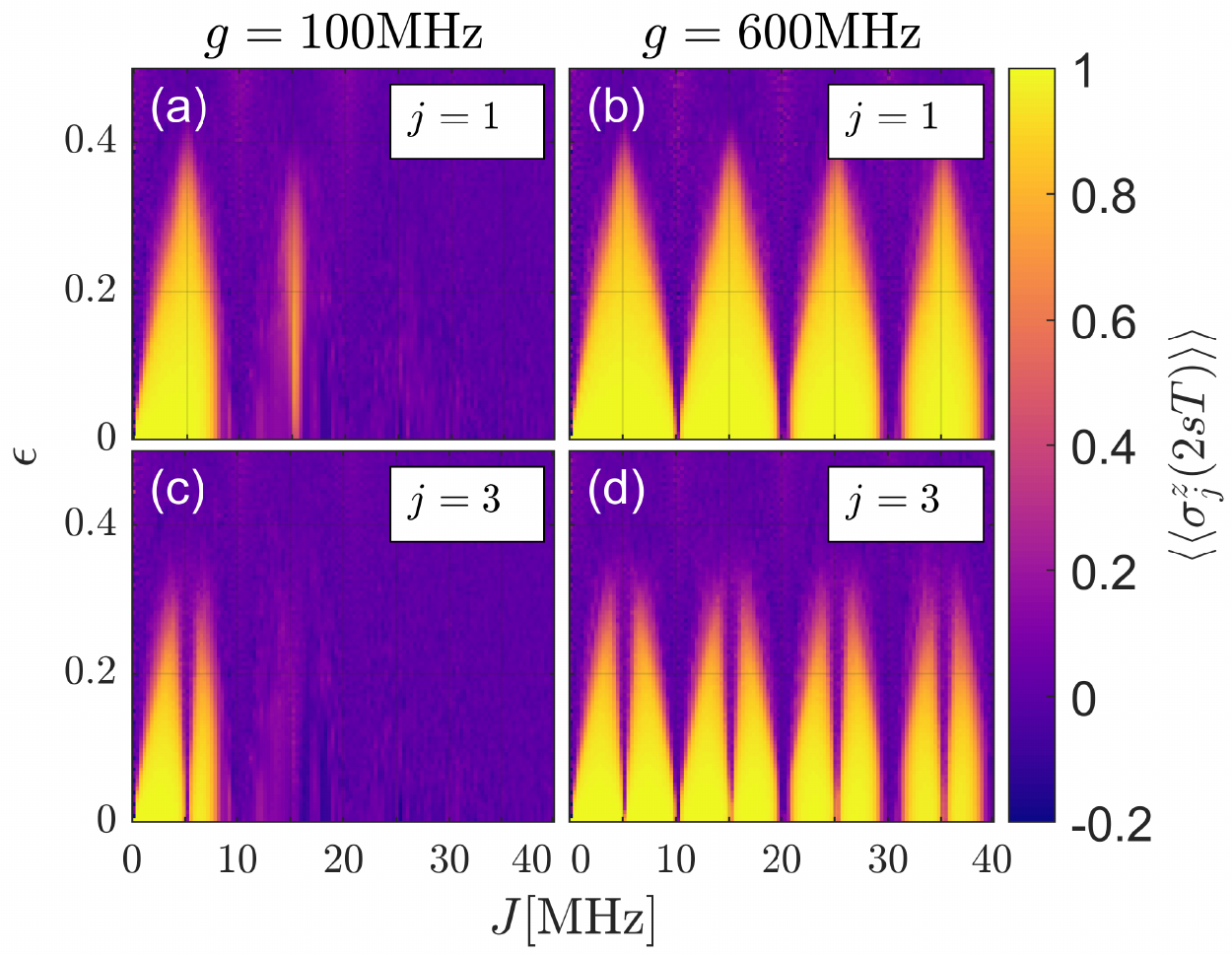}
		\caption{$\epsilon-J$ phase diagram for an $L=6$ spin chain driven by delta-function pulses for gradient strengths of $g=100$ MHz (a,c) and $g=600$ MHz (b,d). The time and disorder-averaged $z$-projections of an end spin ($j=1$, (a,b)) and a bulk spin ($j=3$, (c,d)) are shown. The DTC occurs in the regions where $\langle \langle \sigma^z_1 \rangle \rangle \approx 1$. The initial state is $| \uparrow \downarrow\uparrow \downarrow \uparrow \downarrow \rangle$, and the results are averaged over 100 disorder realizations. The disorder strength and external magnetic fields are $\sigma_B = 9$ MHz and $B_0 = 5$ GHz, respectively, and the driving period is $T=100$ ns.
			\label{fig:PDdeltapulsegrad}}
\end{figure}

Fig.~\ref{fig:PDdeltapulsegrad} shows $\langle \langle \sigma^z_j(2sT) \rangle \rangle$ as a function of the pulse error $\epsilon$ and Heisenberg coupling strength $J$ for end ($j=1$) and bulk ($j=3$) spins of an $L=6$ spin chain for two values of the field gradient $g$. Both the end and bulk spins remain close to the their initial states at late times over a range of pulse errors and coupling strengths; we interpret this is a signature of time crystalline behavior. It is evident from the figure that the end and bulk spins exhibit slightly different behavior as a function of $J$. We discuss the origin of this difference in more detail below. We can view the plot for the end spin as an effective phase diagram and identify the DTC phase as the region in which $\langle\langle \sigma^z_1 \rangle\rangle$ stays close to its initial value ($\langle\langle \sigma^z_1(0) \rangle\rangle=1$ in this case). As was shown in Ref.~\cite{Barnes2019}, the value of  $\langle \langle \sigma^z_1 \rangle \rangle$ remains close to zero when $g=0$.  In contrast, here we see that $g=100$ MHz yields a robust DTC for small, nonzero $J$ and moderate values of $\epsilon$ [Fig.~\ref{fig:PDdeltapulsegrad}(a)].  Increasing $g$ stabilizes the DTC for progressively larger values of $J$ [Fig.~\ref{fig:PDdeltapulsegrad}(b)].  

The stabilization of the DTC phase with increasing gradient field can be understood using perturbation theory.  Consider the following Schrieffer-Wolff transformation:
	\begin{equation}\label{eq:SW}
	S^{(1)}=-J\sum_{j=1}^{L-1} \frac{\sigma_j^+\sigma_{j+1}^- - \sigma_j^-\sigma_{j+1}^+}{2\Delta_{j,j+1} - J(\sigma_{j-1}^z - \sigma_{j+2}^z)}.
	\end{equation}
 In the above expression, we define $\sigma^z_{0} \equiv \sigma^z_{L+1} \equiv 0$, and $\Delta_{j,j+1} = B_{j+1} - B_{j}\approx g$. After this transformation, we obtain $e^{S^{(1)}} \mathcal{H}_H e^{-S^{(1)}} = \mathcal{H}_I + O(J^2/g)$, where $\mathcal{H}_I$ is an Ising model with a gradient field:
	\begin{equation}\label{eq:HIsing}
	\begin{aligned}
	\mathcal{H}_I = \frac{J}{4} \sum_{j=1}^{L} \sigma^z_j \sigma^z_{j+1} + \sum_{j=1}^{L} \frac{B_j}{2} \sigma^z_j.
	\end{aligned}
	\end{equation}
Thus, we see that in the limit $J/g \ll 1$, the gradient-field Heisenberg model becomes an effective Ising model, for which the DTC is stable, as shown in prior work \cite{Else2016,Yao2017,Barnes2019}. In the Appendix, we derive the leading-order corrections to the transformed Hamiltonian and verify the validity of perturbation theory numerically. The fact that these corrections scale like $J/g$ explains why the $\epsilon-J$ phase diagram in Fig.~\ref{fig:PDdeltapulsegrad} ``fills in" from left to right as $g$ is increased.

The gradient therefore plays a similar role as the H2I pulse sequence introduced in Ref.~\cite{Barnes2019}. We find that in order to obtain the same degree of stabilization as found with a gradient of $g=600$ MHz, one would need roughly 300 H2I pulses per period, which would be quite challenging to implement experimentally given current pulse generator bandwidth limitations. Thus, the use of strong magnetic field gradients offers a more practical route to obtain DTCs in exchange-coupled QD spin arrays. On the other hand, an advantage of the H2I approach is that one may dynamically choose which axis to preserve spin states along by choosing the pulse rotation axis appropriately, whereas gradients induced by micromagnets are fixed \cite{Barnes2019}. One may consider how spin state preservation depends on the direction of the magnetic gradient.  To this end, decompose the total field as $\vec{B}_{tot} = \vec{B}_1 + \vec{B}_g$, where $\vec{B}_1$ represents the applied field plus random disorder along the same axis, while $\vec{B}_g$ is a linear field gradient.  We have shown above that when $\vec{B}_1 \parallel \vec{B}_g$, the projection of an end spin along this direction is generically well-preserved in the DTC phase.  On the other hand, we find that products of transverse spin states, such as the $\sigma^y_j$ eigenstate $| \uparrow_y \uparrow_y \uparrow_y \uparrow_y \rangle$, are not at all preserved for $\vec{B}_1 \parallel \vec{B}_g \parallel \hat{z}$, which is consistent with the H2I pulse results from Ref.~\cite{Barnes2019}. Interestingly, we also find that if the gradient term is transverse, i.e., $\vec{B}_1 \perp \vec{B}_g \parallel \hat{y}$, eigenstates of $\sigma^y_j$ are also not preserved.  Thus, it appears to be crucial that the field gradient be parallel to the applied field in order to preserve spin states.

We now investigate the origin of the quasi-periodic patterns evident in Fig.~\ref{fig:PDdeltapulsegrad} and what causes the end spins to behave differently from spins in the middle of the chain. To simplify the analysis, we consider delta-pulse driving, and we focus on the large $g/J$ limit, which allows us to treat the problem using the effective Ising Hamiltonian, ${\cal H}_I$, given in Eq.~\eqref{eq:HIsing}. If the pulse error $\epsilon$ is small, then the periodic $\pi$ pulses perform dynamical decoupling, which effectively removes the local magnetic field terms from the Hamiltonian. This is captured by the stroboscopic Hamiltonian, ${\cal H}_{eff}$, which generates the effective evolution course-grained over two driving periods. To derive ${\cal H}_{eff}$, first consider the exact evolution operator over two periods:
\begin{eqnarray}
    U(2T)&=&R_x(\pi-2\epsilon) e^{-i{\cal H}_IT}R_x(\pi-2\epsilon) e^{-i{\cal H}_IT}\nonumber\\ &=&(-1)^LR_x(-2\epsilon)e^{-i\overline{\cal H}_IT}R_x(-2\epsilon)e^{-i{\cal H}_IT}\nonumber\\ &=&(-1)^LR_x(-2\epsilon)e^{i\epsilon{\cal X}}e^{-i\tfrac{JT}{2}\sum_j\sigma_j^z\sigma_{j+1}^z}\nonumber\\ &\approx& (-1)^Le^{i\epsilon(\sum_j\sigma_j^x+{\cal X})}e^{-i\tfrac{JT}{2}\sum_j\sigma_j^z\sigma_{j+1}^z},\label{U2T}
\end{eqnarray}
where $R_x(\phi)$ denotes an $x$-rotation by angle $\phi$ applied to all spins, $\overline{\cal H}_I=\tfrac{J}{4}\sum_j\sigma_j^z\sigma_{j+1}^z-\sum_j\tfrac{B_j}{2}\sigma_j^z$, and ${\cal X}=\sum_j\sigma_j^x+[-i\overline{\cal H}_IT,\sum_j\sigma_j^x]+\tfrac{1}{2!}[-i\overline{\cal H}_IT,[-i\overline{\cal H}_IT,\sum_j\sigma_j^x]]+\ldots$ is a time-independent operator that arises from commuting $R_x(-2\epsilon)$ past $e^{-i{\cal H}_IT}$. In the last line of Eq.~\eqref{U2T}, we neglected terms of order $\epsilon^2$ in the exponent. It is clear from the final expression that in the absence of pulse errors, the stroboscopic Hamiltonian is $\tfrac{J}{4}\sum_j\sigma_j^z\sigma_{j+1}^z$, which reflects the effect of dynamical decoupling. The presence of a nonzero pulse error generates an additional instantaneous term that slightly rotates all the spins once every two periods, so that the stroboscopic Hamiltonian becomes
\begin{equation}
    {\cal H}_{eff}=\frac{J}{4}\sum_j\sigma_j^z\sigma_{j+1}^z-\epsilon\sum_{s=1}^\infty\delta(t-2sT)\Big(\sum_j\sigma_j^x+{\cal X}\Big).
\end{equation}
In the limit of small $\epsilon$, we can employ first-order perturbation theory to obtain the probability, $P(t)$, to transition from the initial state $|\psi_i\rangle$ to a final state $|\psi_f\rangle$, both of which we assume are eigenstates of $\tfrac{J}{4}\sum_j\sigma_j^z\sigma_{j+1}^z$:
\begin{equation}
    P(t)=\epsilon^2\bigg|\sum_{s=1}^\infty e^{-i2sT\Delta E}\theta(t-2sT)\langle\psi_f|\Big(\sum_j\sigma_j^x+{\cal X}\Big)|\psi_i\rangle\bigg|^2,
\end{equation}
where $\Delta E=E_i-E_f$ is the energy difference of the two states. At late times $t\gg T$, a large number of terms in the sum over $s$ are retained, and these terms will add up incoherently unless $2T\Delta E=2\pi n$, where $n$ is an integer. If the two states differ by the flipping of an end spin, then $\Delta E=\pm J/2$, and we find that $P(t)$ is significant when $JT=2\pi n$. For $T=100$ ns, this corresponds to $J=2\pi n\times10$ MHz, which is consistent with Fig.~\ref{fig:PDdeltapulsegrad}(b). On the other hand, for bulk spins we have $\Delta E=\pm J$, which means transitions out of the initial state are most likely when $JT=\pi n$, or $J=2\pi n\times5$ MHz, which agrees with Fig.~\ref{fig:PDdeltapulsegrad}(d).
	
	\begin{figure}
		\includegraphics[scale=0.62]{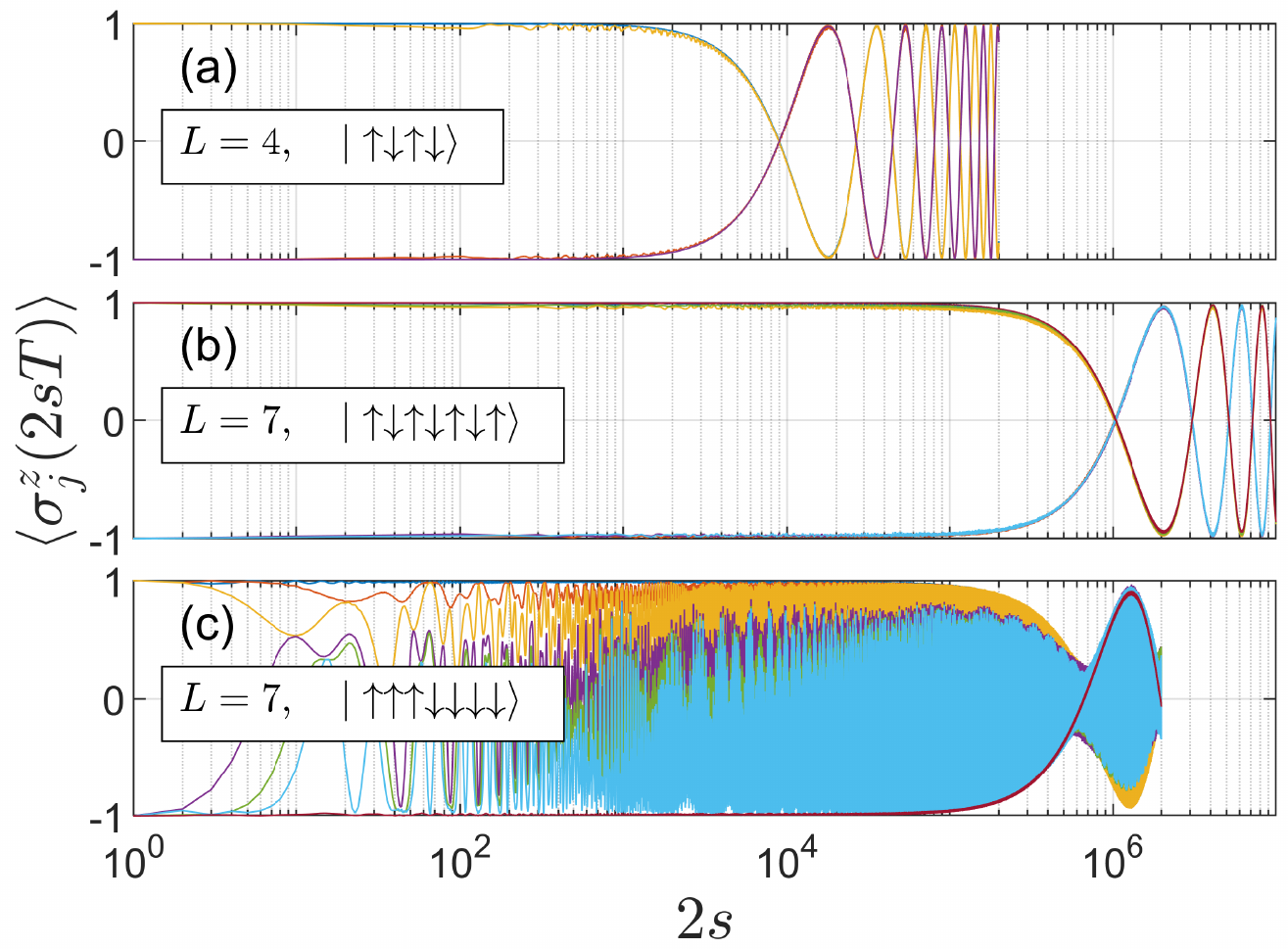}
		\caption{Long-time behavior of $\avg{\sigma_j^z(t)}$ under delta-pulse driving. Only the values after every two driving periods are shown ($t=2sT$ for $s\in\mathbb{Z}^+$).  (a) $L=4$ chain with initial N\'{e}el state. (b) $L=7$ chain with initial N\'{e}el state. (c) $L=7$ chain with initial state $|\uparrow\uparrow\uparrow\downarrow\downarrow\downarrow\downarrow\rangle$. Different colors denote different spins. In (c), only the end spins preserve their initial values. Other parameters are $J = 2.5$ MHz, $T=100$ ns, $g=600$ MHz, $\sigma_B = 9$ MHz, $B_0=5$ GHz, and $\epsilon = 0.1$.}
		\label{fig:lgtRdyndelta}
	\end{figure}
	\begin{figure}
		\includegraphics[scale=0.65]{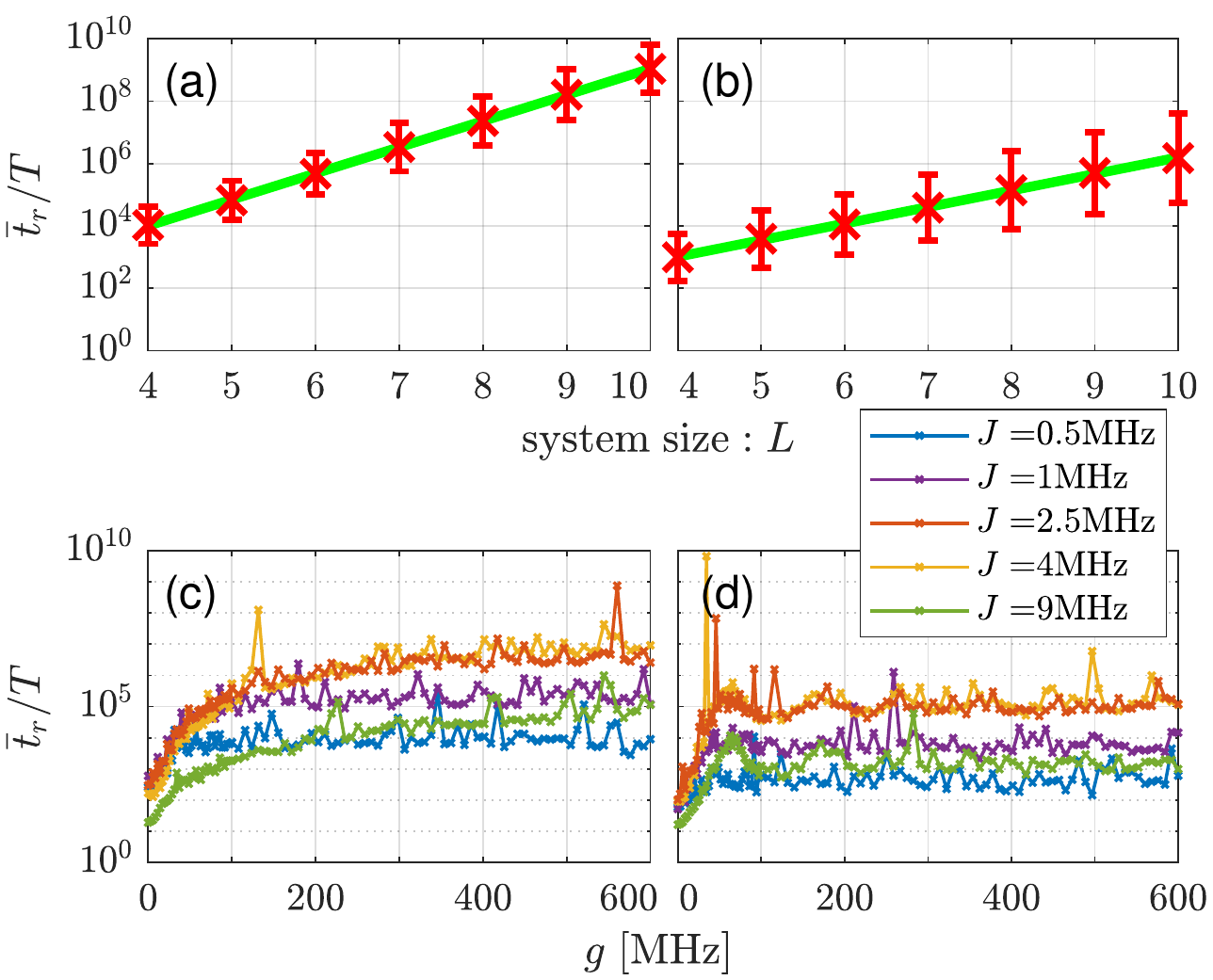}
		\caption{(a,b) Average spin-reversal time $\overline{t}_r$ as a function of system size $L$ for (a) delta-pulse driving and (b) EDSR driving.
		Red crosses indicate numerical results for the spin-reversal times, with the bars giving the standard deviation.  The green lines are exponential best fits. $J=2.5$ MHz, $g=600$ MHz, and results are averaged over 200 disorder realizations. (c,d) Average spin-reversal time for $L=6$ spins as a function of gradient $g$ for various couplings $J$ and for (c) delta-pulse driving and (d) EDSR driving. Results were averaged over 1000 disorder realizations.
		In all panels, the initial state is a N\'{e}el state, while the remaining system parameters are $T = 100$ ns, $\sigma_B = 9$ MHz, $B_0=5$ GHz, $\epsilon=0.1$.}
		\label{fig:lgtR}
	\end{figure}

More detailed information about the spin dynamics can be obtained by tracking the spin expectation values at each site over time for a single disorder realization. Fig. \ref{fig:lgtRdyndelta} shows the time-dependent spin expectation value $\avg{\sigma^z_j (t)}\equiv \Bra{\psi_i} \sigma^z_j (t) \Ket{\psi_i}$ of each site in the chain when it is tuned to the middle of the DTC phase. It is evident in Figs.~\ref{fig:lgtRdyndelta}(a,b) that for an initial N\'eel state, the spin state at each site is preserved for very long times, even up to hundreds of thousands of periods. On the other hand, we see in Fig.~\ref{fig:lgtRdyndelta}(c) that for the initial state $\Ket{\uparrow\uparrow\uparrow\downarrow\downarrow\downarrow\downarrow}$, only the end spins remain frozen in their initial state. This is a consequence of having uniform couplings in Eq.~\eqref{eq:heisenberg} and will be discussed in more detail below when we investigate the mutual information. For now, we focus more on the preservation of the end spins. Fig.~\ref{fig:lgtRdyndelta} shows that this preservation continues up to a timescale $t_r$ when $\langle \sigma^z_1 (t) \rangle$ changes sign; we refer to $t_r$ as the ``spin-reversal time". Beyond this timescale, the $z$-projection of each spin starts to oscillate between $+1$ and $-1$ in such a way that the even sites are perfectly antisynchronized with the odd sites. This can be understood from the fact that $\otimes_{j=1}^L\sigma^x_j$ commutes with the static Ising part of $\mathcal{H}_{eff}$, so that $\Ket{\psi_i}$ and $\otimes_{j=1}^L\sigma^x_j\Ket{\psi_i}$ are degenerate. The driving term then induces slow transitions between these two states at the $L$th order of perturbation theory. This then leads to an exponential growth of the disorder-averaged spin-reversal time, $\overline{t}_r$, with system size [see Fig.~\ref{fig:lgtR}(a)], indicating the importance of many-body interactions in preserving the spin. The spin-reversal time $\overline{t}_r$ also grows with the field gradient as expected given its role in stabilizing the DTC [Fig. \ref{fig:lgtR}(c)]. Without a gradient, $\overline{t}_r$ remains small, whereas even a relatively small gradient drastically improves the spin state preservation time. This occurs for generic interaction strengths, although the benefit varies nonmonotonically as $J$ increases, which is consistent with the phase diagram shown in Fig.~\ref{fig:PDdeltapulsegrad}. Here we have treated the disorder as quasistatic, but in a QD system it is mostly due to the fluctuations of the nuclear spins, and so it varies slowly over timescales longer than a few microseconds \cite{Medford2012,Malinowski2017}. Thus, the true limits on the spin preservation time are likely shorter than the ones obtained within the present model.  Nevertheless, we expect a significant effect to be observable in experiments.

\begin{figure}
		\centering
		\includegraphics[scale=0.65]{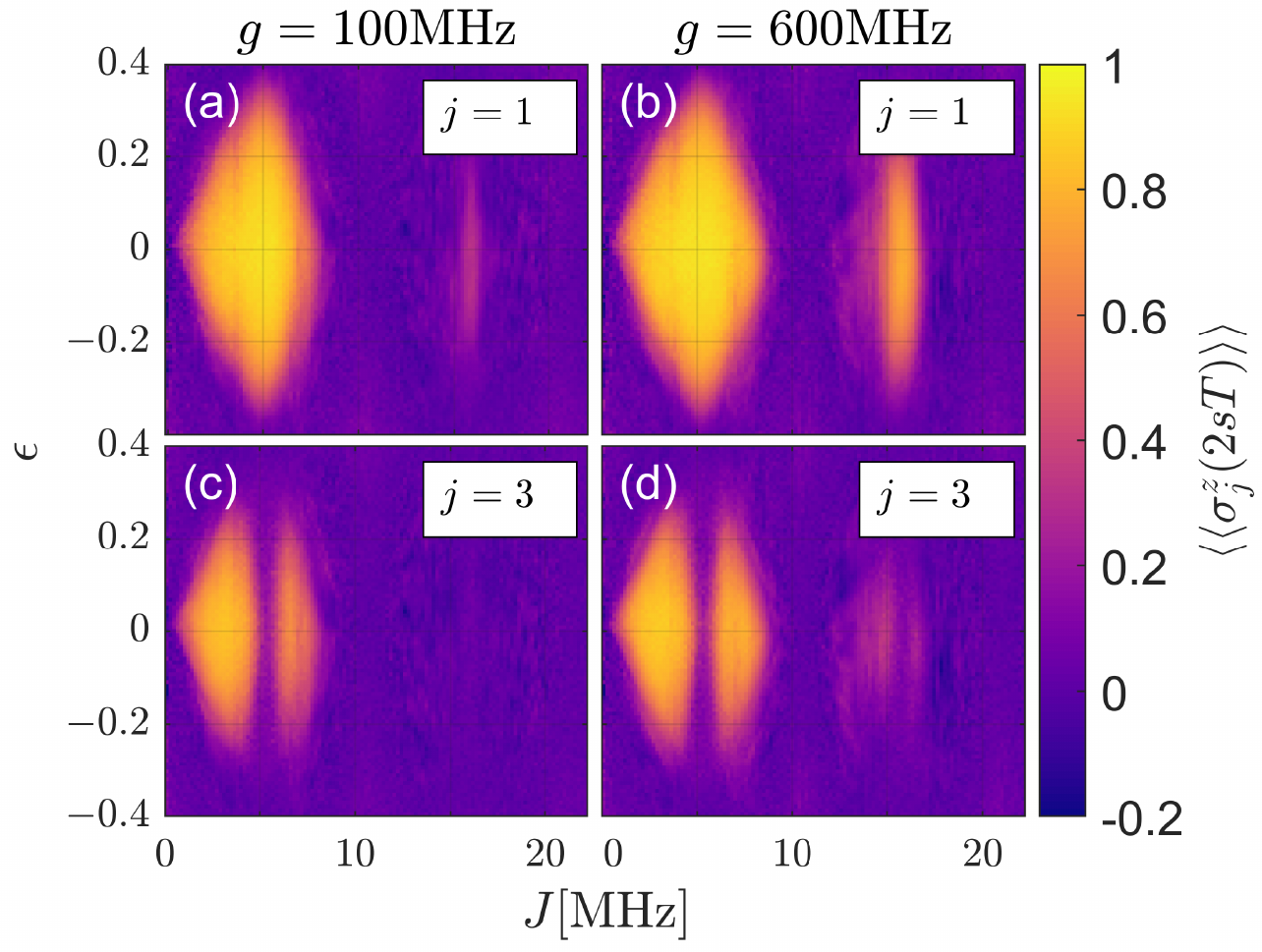}
		\caption{$\epsilon-J$ phase diagram for an $L=6$ spin chain driven by EDSR pulses for gradient strengths of $g=100$ MHz (a,c) and $g=600$ MHz (b,d). The time and disorder-averaged $z$-projections of an end spin ($j=1$, (a,b)) and a bulk spin ($j=3$, (c,d)) are shown. The initial state is $| \uparrow \downarrow\uparrow \downarrow \uparrow \downarrow \rangle$, and the results are averaged over 100 disorder realizations. The pulse duty cycle is $\eta=0.1$, so that the pulse duration is $\eta T=10$ ns. The remaining parameters are as in Fig.~\ref{fig:PDdeltapulsegrad}.}
		\label{fig:pdedsrgrad}
	\end{figure}

\begin{figure}
		\includegraphics[scale=0.56]{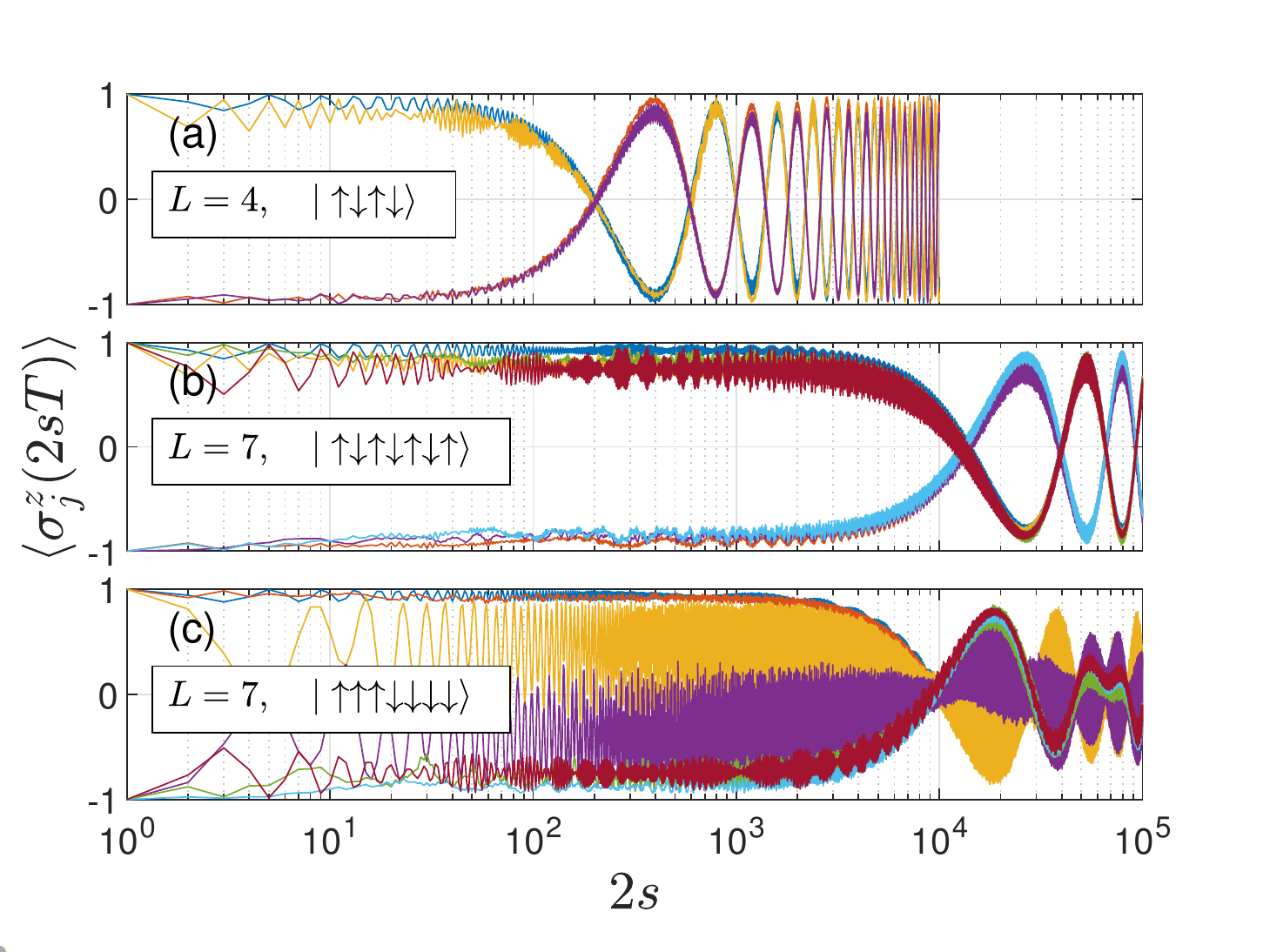}
		\caption{Long-time behavior of $\avg{\sigma_j^z(t)}$ under EDSR driving. Only the values after every two driving periods are shown ($t=2sT$ for $s\in\mathbb{Z}^+$).  (a) $L=4$ chain with initial N\'{e}el state. (b) $L=7$ chain with initial N\'{e}el state. (c) $L=7$ chain with initial state $|\uparrow\uparrow\uparrow\downarrow\downarrow\downarrow\downarrow\rangle$. Different colors denote different spins. In (c), only the edge spins preserve their initial values. $\eta=0.1$, and all other parameters are as in Fig.~\ref{fig:lgtRdyndelta}.}
		\label{fig:lgtRdynEDSR}
	\end{figure}

Next, we consider how the spin state preservation properties are affected when EDSR pulses are used instead of delta-function pulses. While the only source of error associated with the delta-function pulses is the rotation error $\epsilon$, additional errors arise for EDSR pulses. Part of this is due to detuning fluctuations away from resonance caused by the local magnetic field disorder. This should enhance the role of disorder and reduce the state preservation effect in the EDSR case. This is evident in Figs.~\ref{fig:lgtR}(b,d), which show that the enhancement of $\overline{t}_r$ with system size and gradient is more modest compared to delta-pulse driving. To further understand the impact of EDSR driving, we also show the $\epsilon-J$ phase diagram in Fig.~\ref{fig:pdedsrgrad}. In addition to disorder, we also expect the finite pulse duration, $\eta T$, to have a negative impact on state preservation. In particular, when $J\eta T\sim 1$, the spin-spin interactions will interfere with the EDSR pulses and prevent them from completely flipping the spins. For the simulation results shown in Fig.~\ref{fig:pdedsrgrad}, we set $\eta = 0.1$ and $T = 100$ ns, resulting in a driving amplitude of $\frac{\pi}{2}(\eta T)^{-1}= 25$ MHz when $\epsilon=0$. Thus, this interference between interactions and pulses should become severe as $J$ approaches 25 MHz. The figure indeed shows that the state preservation is no longer visible when $J\gtrsim20$ MHz. Also, compared to Fig.~\ref{fig:PDdeltapulsegrad}, the quasi-periodic dips in the average spin projection as a function of $J$ are slightly shifted due to the fact that the free evolution time, $(1-\eta)T$, is effectively reduced in this case. In Fig.~\ref{fig:pdedsrgrad}, we show both positive and negative values of $\epsilon$ because the phase diagram differs slightly in these two cases. This is because the pulse amplitude, $(\frac{\pi}{2}-\epsilon)(\eta T)^{-1}$, depends on the sign of $\epsilon$ in the EDSR case. The impact of EDSR driving for different sites and initial states is illustrated in Fig.~\ref{fig:lgtRdynEDSR}, which shows qualitatively similar behavior to the delta-driving case, Fig~\ref{fig:lgtRdyndelta}, albeit with less perfect and shorter-lived state preservation.
	

\begin{figure}
		\includegraphics[scale=0.5]{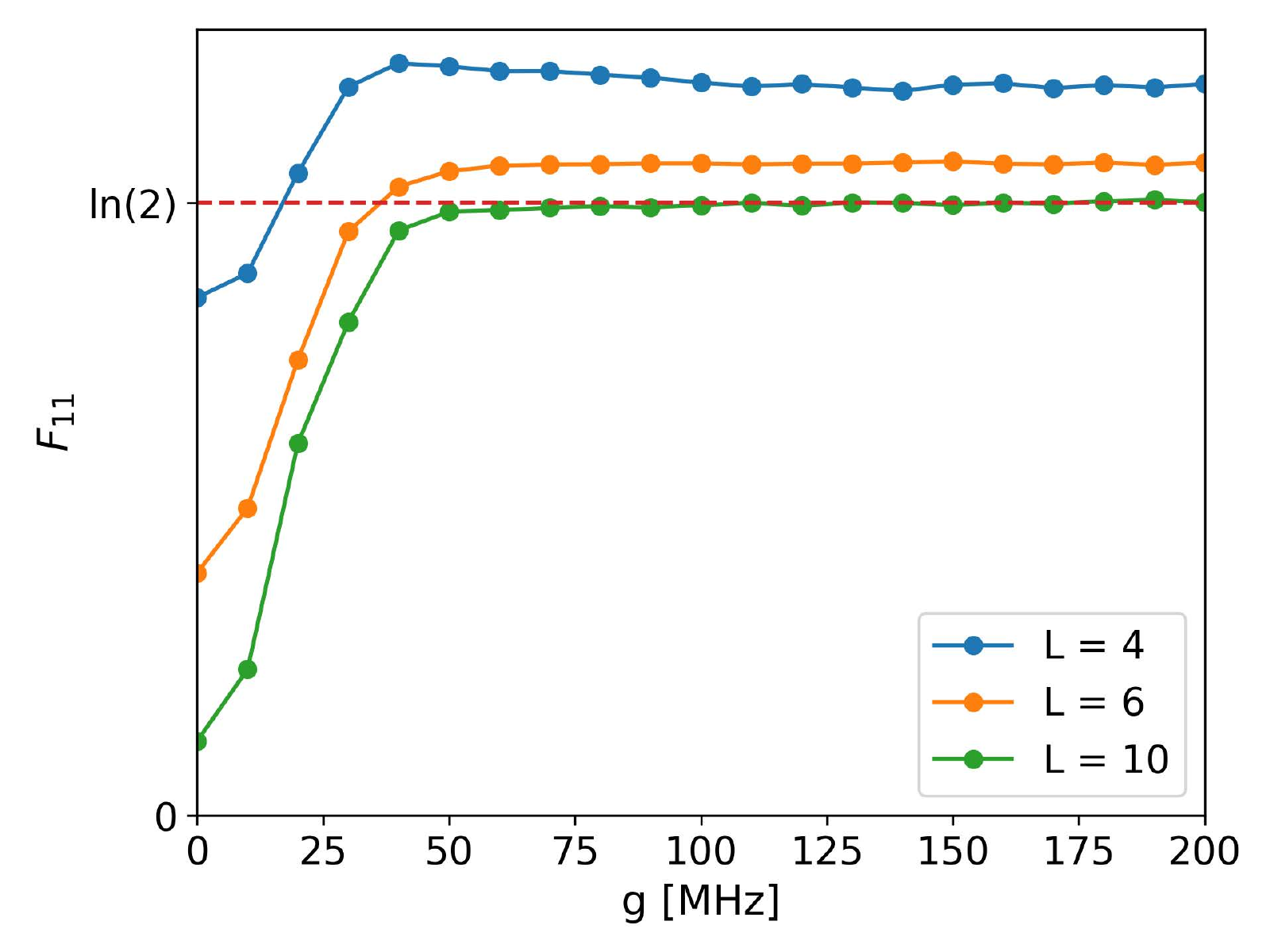}
		\caption{Mutual information $F_{11}$ for $L=4,6,10$ as a function of field gradient.  Other parameters are $B_0 = 5.6$ GHz, $\sigma_B = 9$ MHz, $T=100$ ns, $\epsilon=0.05$, $J=4$ MHz.  Results are averaged over 1000 ($L=4$), 800 ($L=6$), and 20 ($L=10$) disorder realizations. 
			\label{fig:mutualinfo2}}
	\end{figure}

Long-range correlations between end spins are also preserved in the DTC phase. To show this, we calculate the mutual information between the two spins at opposite ends of the chain. In general, the mutual information is given by $I(A,B) = S(A) + S(B) - S(A \cup  B)$, where $S(R) = -\Tr \rho_R \ln \rho_R$ is the von Neumann entropy for the reduced density matrix $\rho_R$ of region $R$.  We calculate this quantity for the Floquet eigenstates of the delta-pulse driven system. In the thermodynamic limit (and also taking the sizes of $A$ and $B$ and their separation distance to infinity), $I(A,B) \rightarrow \ln k$, with $kT$ equal to the period of the observable $\mathcal{O}$ whose expectation value breaks the time translation symmetry of the drive \cite{Else2016}. The integer $k$ is also equal to the number of different terms appearing in a Floquet eigenstate. In Fig.~\ref{fig:mutualinfo2} we show the mutual information $F_{11}$, where the subscripts indicate that the size of the subregions $A$ and $B$ include one site each, located at opposite ends of the chain.  For $L=4$, the mutual information does not converge to the expected value of $\ln(2)$. Nevertheless, the failure of $F_{11}$ to precisely converge to $\ln(2)$ is merely a finite-size effect, as $F_{11}$ quickly approaches the expected value for larger system sizes. We also note that, because we consider a uniform-coupling model (unlike in Ref.~\cite{Else2016}, which included disorder in $J$), the mutual information will generally not converge to $\ln(2)$ for pairs of bulk spins. Equivalently, spin projections in the bulk will not necessarily exhibit the same $2T$ periodicity for every short-range correlated initial state. An example of this was already shown in Fig.~\ref{fig:lgtRdyndelta}(c), where bulk spins do not exhibit $2T$ periodicity for an initial state of the form $\Ket{\uparrow\uparrow\uparrow\downarrow\downarrow\downarrow\downarrow}$. This happens because the uniform couplings lead to a large degeneracy among Floquet eigenstates in the limit of vanishing pulse error. For perturbatively small pulse errors, this then implies that the Floquet eigenstates are superpositions containing more than two terms. However, in a real system such as a QD spin array, some variation in the couplings is inevitable, and we expect the $2T$ periodicity to arise for bulk spin observables as well.
	
	\section{Many-body Localization \label{sec:MBL}}
	
	We now investigate MBL in the \textit{undriven} model, Eq.~\eqref{eq:heisenberg}, looking at diagnostics that are complementary to those studied in recent works \cite{Schulz2019, vanNieuwenburg2019, Wu2019, Taylor2019}. One of our motivations is to confirm that it is indeed MBL and not another type of non-ergodic phase which is responsible for the appearance of a DTC in the driven case. We also reconsider previously studied signatures, with an emphasis on parameter regimes relevant to QD experiments.
	
	The quantum Fisher information (QFI) is a promising tool for characterizing entanglement spreading in many-body systems \cite{Braunstein1994,Smith2016}.  Unlike many other measures (e.g. entanglement entropy), it is directly accessible in experiments.  Recently, the logarithmic growth of the QFI in time was used in the trapped ion experiment of Ref. \cite{Smith2016} to argue for the existence of MBL in that system.  While the general definition of the QFI is somewhat involved, it takes a simple form in the case of pure states.  Given a Hermitian operator $\mathcal{O}$, the (normalized) QFI in a pure state $\psi$ is proportional to the variance of $\mathcal{O}$ in that state: $f_Q(\mathcal{O},\psi) =  (\Delta \mathcal{O})^2/L = (\langle \mathcal{O}^2 \rangle - \langle \mathcal{O} \rangle^2)/L$.  The QFI can be interpreted as a measure of the sensitivity of $\psi$ to the unitary transformation $e^{i \theta \mathcal{O}}$, which is an important quantity in quantum metrology \cite{Smith2016}.  Furthermore, the normalized QFI is an entanglement witness when $f_Q > 1$.  Given that the Heisenberg interaction in our model is antiferromagnetic, we choose $\mathcal{O}$ to be the staggered magnetization, $\mathcal{O} =  \sum_j (-1)^j \sigma^z_j$.
	
	\begin{figure}
		\includegraphics[scale=0.57]{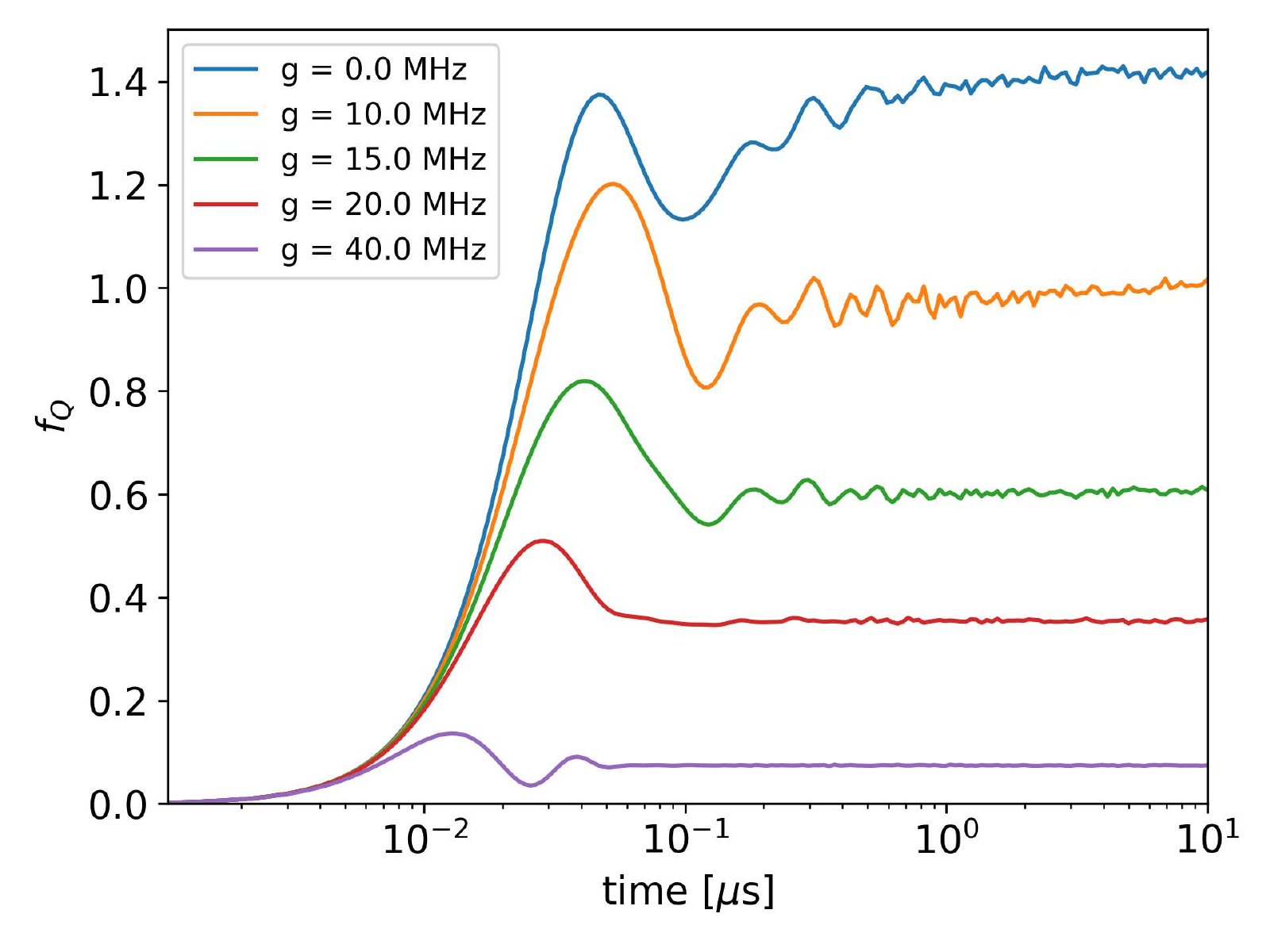}
		\caption{Quantum Fisher information as a function of time and for various values of the magnetic field gradient $g$ (in MHz). Parameters are $L=8$, $J=4$ MHz, $B_0=100$ MHz, $\sigma_B = 4$ MHz.  The initial state is a N\'{e}el state, and results are averaged over 1800 disorder realizations. \label{fig:QFIvst}}
	\end{figure}
	
	\begin{figure}
		\includegraphics[scale=0.62]{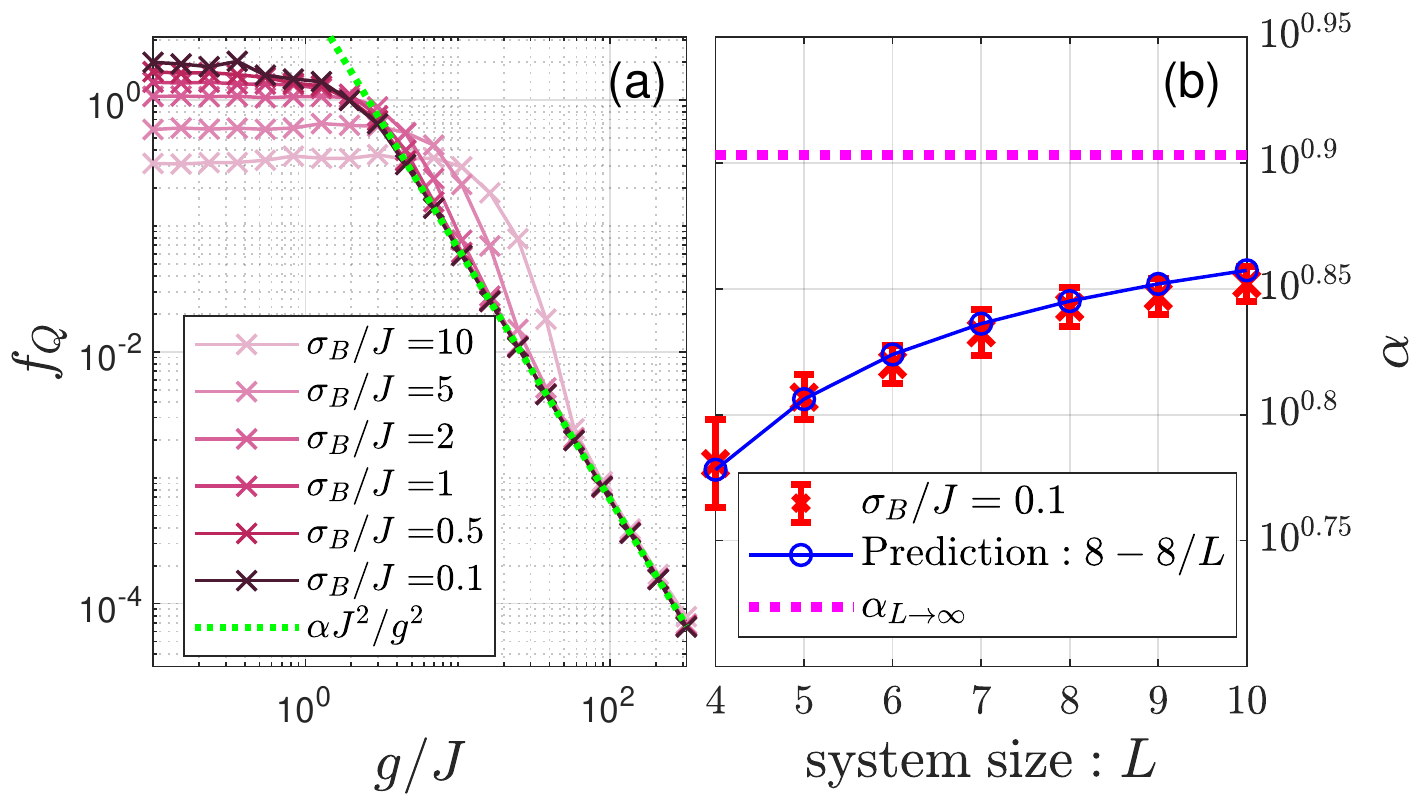}
		\includegraphics[scale=0.7]{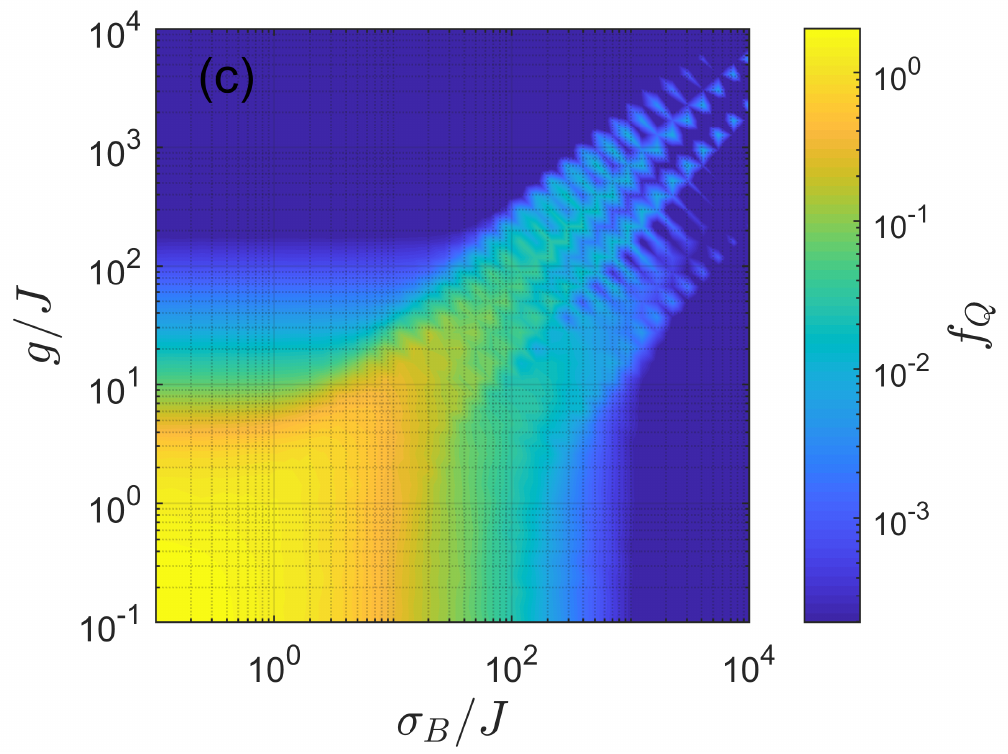}
		\caption{Late-time behavior of the QFI as a function of gradient and disorder strength. (a) QFI $f_Q$ as a function of gradient $g$ for different disorder strengths $\sigma_B$. The results at large $g$ fall onto a straight line corresponding to $\alpha J^2/g^2$. (b) The value of $\alpha$ for different system sizes $L$ obtained from numerical fitting (red points) and from perturbation theory (blue circles). The value of $\alpha$ in the thermodynamic limit ($\alpha=8$) is also shown (dashed line). (c) QFI $f_Q$ as a function of gradient $g$ and disorder strength $\sigma_B$. Here, we set $B_0=5\times10^3J$. Each point was time-averaged over the range $10^3<Jt<10^4$ by sampling 24 times within this duration for each disorder realization. 50 disorder realizations were averaged over for each data point in the plot.  \label{fig:QFIlatetime}}
	\end{figure}

In Fig. \ref{fig:QFIvst} we present the normalized disorder-averaged QFI as a function of time for different field gradients.  After an initial transient regime, the disorder-averaged QFI becomes constant out to late times (individual realizations show oscillations around a mean value).  For small $g/J$, the system is in the ergodic phase, characterized by growth of the QFI at intermediate times and a large late-time QFI, indicating the presence of entanglement.  As $g/J$ increases, the overall magnitude of the QFI is highly suppressed, suggesting a decrease in the entanglement as one enters the MBL phase.  (One cannot claim this with certainty, as the QFI provides only a sufficient, not a necessary, condition for entanglement.) This suppression of the late-time QFI with increasing gradient is analyzed in more detail in Fig.~\ref{fig:QFIlatetime}. In Fig. \ref{fig:QFIlatetime}(a), we show that at as $g$ becomes large, the QFI decays as a power law, $f_Q\approx\alpha(J/g)^2$, regardless of the amount of disorder, whereas it approaches a disorder-dependent constant value for small $g$. The origin of the large-$g$ power-law behavior can be understood using the same Schrieffer-Wolff transformation as before [Eq.~\eqref{eq:SW}], but this time applied on $\sigma_j^z$:
\begin{equation}\label{Appeq:Psi}
	{\cal Z}_j = \sigma_j^z + [S^{(1)},\sigma_j^z] + \frac{1}{2}[S^{(1)},[S^{(1)},\sigma_j^z]],
	\end{equation}
where we neglect terms of order ${\cal O}(J^3/g^3)$ and beyond. The explicit form of ${\cal Z}_j$ is given in Appendix~\ref{app:QFIpert}. We can use these operators to obtain the late-time QFI by computing expectation values of them with respect to the spin state at late times:
\begin{equation}
	\begin{aligned}
	f_Q &= \frac{1}{L}\sum_{m,n=1}^L\mathcal{N}_m^{-1} \mathcal{N}_n^{-1}(-1)^{m+n}\left(
	\avg{{\cal Z}_m{\cal Z}_n} - \avg{{\cal Z}_m}\avg{{\cal Z}_n}
	\right)\\
	&=
	\left(8-8/L\right)J^2/g^2 + \mathcal{O}(J^4/g^4).
	\end{aligned}\label{eq:fQanalytical}
	\end{equation}
Here, ${\cal N}_j=\sqrt{\avg{{\cal Z}_j^2}}$ are normalization factors. To obtain the final expression in Eq.~\eqref{eq:fQanalytical}, we use the fact that the system remains close to its initial state (here, taken to be the N\'eel state $\Ket{\uparrow\downarrow\uparrow\downarrow\cdots}$) at late times, which in turn allows us to compute the expectation values explicitly. Details can be found in Appendix~\ref{app:QFIpert}. The result of Eq.~\eqref{eq:fQanalytical} implies that the coefficient of the large-$g$ power law is $\alpha=8-8/L$. This agrees well with numerically fitted values of $\alpha$, as shown in Fig.~\ref{fig:QFIlatetime}(b). 

Fig.~\ref{fig:QFIlatetime}(c) shows the QFI as a function of both the gradient and disorder strength. For small $g$ and $\sigma_B$ the system is ergodic, as evidenced by the large value of the QFI, whereas outside this region the QFI is suppressed, indicating localization. In the large $g$ regime, the localization follows from the fact that the interactions are effectively of Ising type [Eq.~\eqref{eq:HIsing}], and thus the initial product state is approximately an energy eigenstate. Localization for $\sigma_B\gtrsim J$ is consistent with what is known about the MBL phase transition in disordered Heisenberg chains \cite{Pal2010,Barnes2016}. Fig.~\ref{fig:QFIlatetime}(c) shows that the gradient has a stronger effect in localizing the system compared with a comparable level of disorder.

Another indication of MBL is the failure of the system to absorb energy from a driving field. To examine this, we first define a dimensionless energy parameter $Q(t)$:
	\begin{equation}
	\begin{aligned}
	Q(t) = \frac{E(t) - E(0)}{E_{\infty}-E(0)},
	\end{aligned}
	\end{equation}
	where $E_0$ is the energy of the initial state and $E_\infty = \Tr [\mathcal{H}]/N_s$ is the energy of the infinite temperature state, where $N_s$ is the number of states in the Hilbert space. As $t \rightarrow \infty$, $Q \rightarrow 1$ if the system absorbs energy and heats up to infinite temperature; $Q \approx 0$ if the system does not heat up. We consider the energy absorption process in the case where the system is driven by a periodic train of square pulses with amplitude $A$, period $T$, and duty cycle $\eta$.  The pulse consists of an additional $B_x$ field applied to the even-numbered sites in the chain: 
	\begin{equation}
	\begin{aligned}
	\mathcal{H}_p = \sum_{k=1}^{\infty} \sum_{j=1}^{L/2} A [\theta (t/T - (k-\eta)) - \theta(t/T - k)] \sigma^x_{2j} .
	\end{aligned}
	\end{equation}
	Because the driving field is applied to every other site in the chain, the commutator of $\mathcal{H}_p$ with the Heisenberg interaction term is maximized, and this in turn accelerates heating in the system through the non-conservation of energy.
	
\begin{figure}
		\includegraphics[scale=0.49]{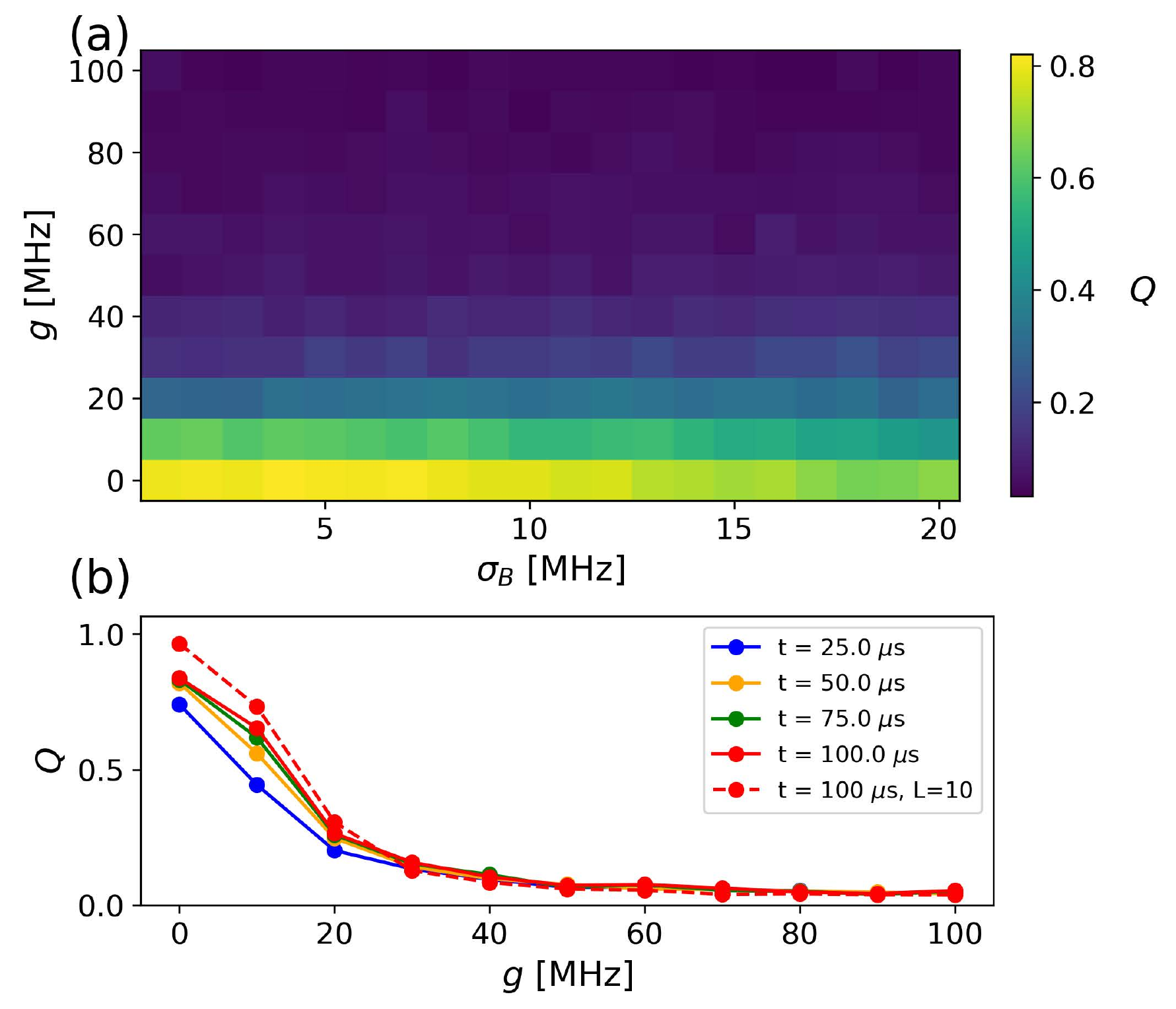}
		\caption{(a) Late-time dimensionless energy $Q$ as a function of gradient $g$ and field disorder strength $\sigma_B$ under periodic driving.  Large $Q$ indicates the heating of the system towards infinite temperature, which occurs in the ergodic phase. For $Q \ll 1$ the system does not absorb energy, indicating an MBL phase.  (b) $Q$ vs. $g$ for $\sigma_B = 5.0$ MHz, evaluated at different times $t$ (in units of the driving period $T$). As $t$ increases, the curves approach a fixed limiting curve $Q(g)<1$, in contrast to the expectation for prethermalization, for which $Q(g)$ would ultimately approach 1 at long times. The system parameters are $L=8$, $J\!=\!4$ MHz, $B_0\!=\!100$ MHz, with the initial state being the ground state. The driving parameters are $A\!=\!50$ MHz, $T\!=\!10$ $\mu$s, $\eta = 0.5$. The dashed red line shows $Q$ for $L=10$, which yields values closer the expected infinite temperature limit. For (a), the $Q$ values are obtained by averaging over the last 500 periods for simulations of length $4000T$. Results are averaged over 96 disorder realizations. \label{fig:heating}}
	\end{figure}

	In Fig.~\ref{fig:heating}(a), we present the energy absorption $Q$ as a function of field disorder $\sigma_B$ and gradient $g$, using parameters relevant for QD arrays. One sees that relatively small $g$ is already sufficient to suppress $Q$, indicating the transition from the ergodic to the localized phase. In contrast, without a  gradient the system remains susceptible to heating, even for longitudinal field disorder significantly larger than what is typically measured in GaAs QDs. It is natural to ask whether the failure of the system to heat up extends out to arbitrarily long times, or if the system eventually heats up on sufficiently long timescales. The latter would indicate the that the localized phase is prethermal rather than MBL. In this case, one might expect that curves of $Q(g)$, plotted for successively later times, would ultimately approach $Q(g) \approx 1$ for all values of $g$. Fig.~ \ref{fig:heating}(b) reveals that this is not case: the $Q$ results approach a fixed curve, with $Q(g) \neq 1$, thus confirming that the gradient-induced localization is indeed MBL and not a prethermal phase of matter. We also note that while the results for an array of size $L=8$ do not achieve values $Q \approx 1$ in the ergodic phase, considering larger system sizes increases the maximum $Q$ that can be achieved (see the dashed red line for $L=10$ in Fig.~ \ref{fig:heating}(b)).
	
\begin{figure}
		\centering
		\includegraphics[scale=0.6]{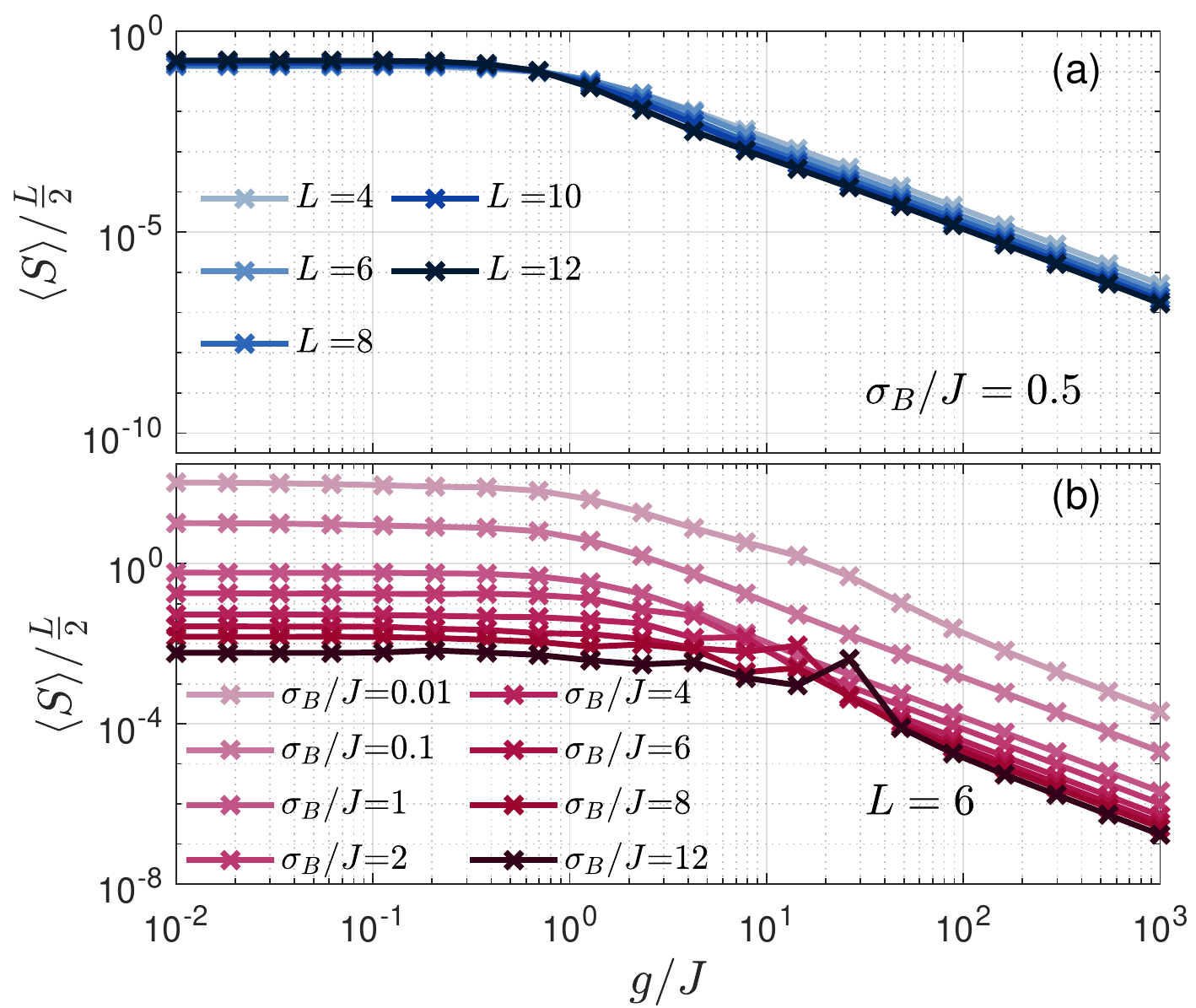}
		\caption{Normalized bipartite entanglement entropy versus gradient-coupling ratio $g/J$ for (a) different system sizes $L$ and (b) various disorder strengths $\sigma_B$. In (a) we set $\sigma_B=0.5J$, while for (b) we set $L=6$. In both panels, we set $B_0/J=5\times 10^3$. Numerical fitting yields a power law $\sim (g/J)^{-\nu},\;\nu=1.84\pm 0.05$ for large $g/J$.}
		\label{fig:EEloglog}
	\end{figure}

	Finally, we examine the entanglement entropy and participation ratio of the undriven Heisenberg spin chain with magnetic field gradient. These quantities have been studied previously for disordered Heisenberg spin chains \cite{Pal2010,Luitz2015,Barnes2016} and for spinless fermions in an electric field \cite{Schulz2019,vanNieuwenburg2019}, the latter being equivalent to a spin chain with a gradient. We make a direct comparison to these earlier works in Appendix~\ref{app:MBLcompare}. Here, we instead focus on the gradient dependence of these diagnostics in parameter regimes relevant for QD experiments. The bipartite entanglement entropy is defined as $S(A) = -\Tr \rho_A  \ln \rho_A$, where $A$ is taken as the left half of the chain. When divided by the system size, $S(A)/L$, this quantity approaches zero as $L \rightarrow \infty$ in the localized phase (the so-called area law), but remains finite if the system is delocalized (volume law). These trends are evident in Fig. \ref{fig:EEloglog}, where we present the entanglement entropy as a function of field gradient for several different system sizes and disorder strengths. As $g$ becomes small, $S(A)/L$ approaches a finite, fixed value as $L \rightarrow \infty$, indicating the volume law expected for an ergodic phase.  On the other hand, at large $g$, $S(A)/L$ goes to zero, as expected for area law entanglement and MBL. We find that this decay is well described by a power law of the form $(g/J)^{-\nu}$, where $\nu=1.84\pm0.05$ is obtained through numerical fitting. 
	
\begin{figure}
		\centering
		\includegraphics[scale=0.6]{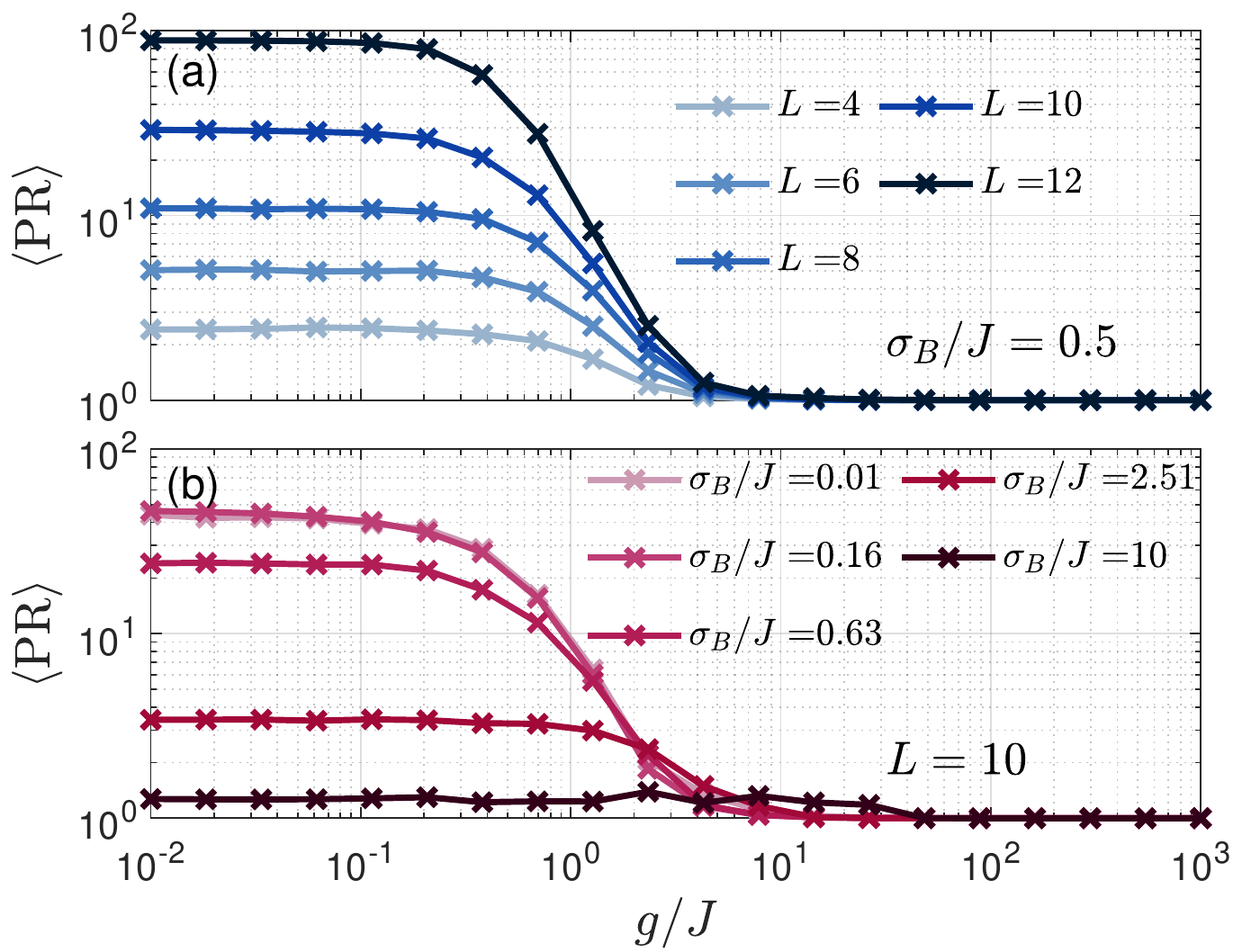}
		\caption{Participation ratio PR as a function of the gradient-coupling ratio $g/J$ for (a) different system sizes $L$ and (b) various disorder strengths $\sigma_B$. In (a) we set $\sigma_B=0.5J$, while for (b) we set $L=10$. In both panels, we set $B_0/J=5\times 10^3$.
		}
		\label{fig:PRloglog}
	\end{figure}

	The participation ratio is defined by $\mathrm{PR}\equiv \avg{(\sum_{k=1}^{2^L}|\psi_{i,k}|^4)^{-1}}$, where $\psi_{i,k}$ is the overlap of the initial state $\Ket{\psi_i}$ with the $k$th energy eigenstate. $\mathrm{PR}$ converges to $1$ in a localized phase because each overlap is approximately given by $|\psi_{i,k}|=\delta_{ik}$. In the ergodic phase, it scales proportionally to the system size. Fig.~\ref{fig:PRloglog} shows that these limiting behaviors occur for large and small $g/J$, respectively, providing further evidence of an ergodic to MBL transition as $g/J$ increases. Both the entanglement entropy and participation ratio indicate that the system is well within the MBL phase for $g\gtrsim 10J$ over a broad range of system sizes and disorder strengths. This is consistent with the behavior of both the QFI and heating diagnostics considered earlier (see Figs.~\ref{fig:QFIlatetime} and \ref{fig:heating}). It is also consistent with the value of $g$ that demarks the onset of a DTC phase in the driven system, as evidenced by the mutual information (Fig.~\ref{fig:mutualinfo2}). The regime $g\gtrsim 10J$ is readily achieved in both Si and GaAs QD experiments using either a micromagnet or nuclear spin programming to create a magnetic field gradient \cite{Kawakami2016,Nichol2017,Sigillito2019}. Gradients on the order of several hundred MHz have been created in this way, while $J$ can typically be tuned via gate voltages from zero up to several hundred MHz or more \cite{Nichol2017,Takumi2018,Kandel2019,Mills2019,Sigillito2019,Dehollain2019,Yang2019,Huang2019,Xue2019,Cerfontaine2019}. Thus, we expect that both MBL and DTC phases should be observable in current QD experiments.


	\section{Conclusions \label{sec:conclusions}}
	We have shown that the recently studied many-body localized phase induced by a magnetic field gradient also hosts a time crystal phase under periodic driving.  The latter is evidenced by long-time preservation of spin states and the asymptotic form of the mutual information.  In particular, this phase is realized in a Heisenberg chain with a gradient field, and is experimentally relevant in quantum dot systems.  Notably, the spin dynamics of edge and inner spins differ as a function of the Heisenberg coupling and pulse error.  Furthermore, alternative diagnostics of many-body localization in the undriven model, such as the quantum Fisher information and energy absorption under driving, broadly agree with the signatures recently studied in the literature. Our results show that many-body localization and time crystal phases should be realizable in quantum dot arrays using only demonstrated capabilities.
	
	\begin{acknowledgments}
	We thank Sriram Ganeshan and John Nichol for helpful discussions. This  work  is  supported  by  DARPA  Grant  No. D18AC00025.
	\end{acknowledgments}
	
	\appendix
	\section{Schrieffer-Wolff Transformation}\label{app:SW}

The Schrieffer-Wolff transformation \cite{Schrieffer1966} for the gradient-field Heisenberg model presented in Eq.~\eqref{eq:SW} illustrates how the presence of a strong magnetic gradient suppresses the transverse spin-flipping terms, producing an effective Ising model at lowest order in perturbation theory.  To see this, write $\mathcal{H}_H = \mathcal{H}_0 + \mathcal{H}_1$, with

\begin{align}
\mathcal{H}_0 &= \frac{J}{4} \sum_{j=1}^{L} \sigma^z_j \sigma^z_{j+1} + \sum_{j=1}^{L} \frac{B_j}{2} \sigma^z_j \\
\mathcal{H}_1 &= \frac{J}{4} \sum_{j=1}^{L} \Big( \sigma^x_j \sigma^x_{j+1} + \sigma^y_j \sigma^y_{j+1} \Big)
\end{align}

	\begin{figure}
		\includegraphics[scale=0.5]{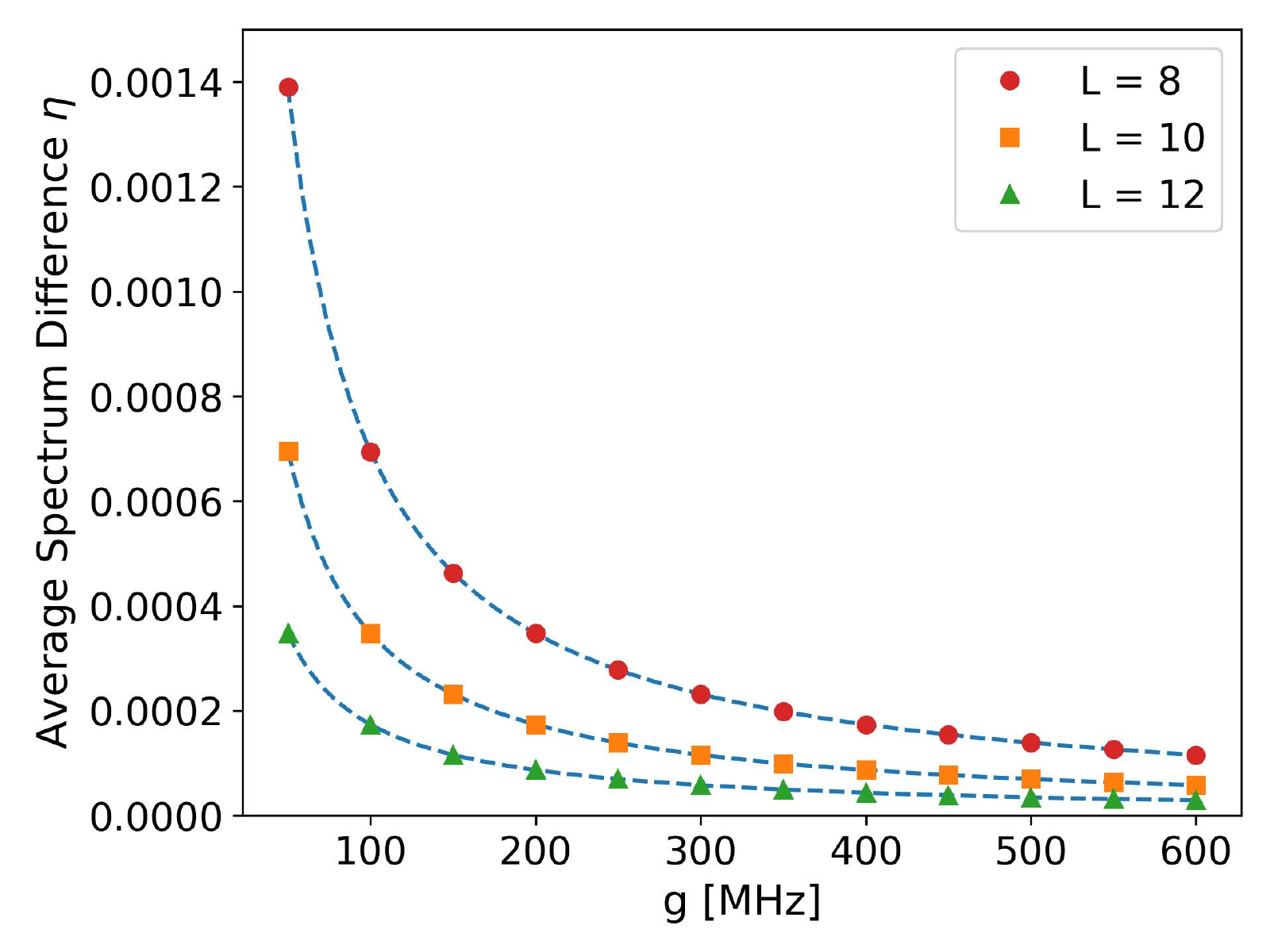}
		\caption{Average energy eigenvalue difference $\eta$ between the Schrieffer-Wolff-transformed Heisenberg model and the corresponding Ising model, as a function of magnetic field gradient, and for several system sizes.  The dashed blue lines are fits to $\propto 1/g$, which is the lowest-order correction to the transformation. Parameters are $J=1$, $B_0=0$, $\sigma_B=0$.  \label{fig:SWtest}}
	\end{figure}

The Schrieffer-Wolff transformation is constructed to eliminate $\mathcal{H}_1$ to lowest order in perturbation theory, $\mathcal{H}_1 + [S^{(1)},\mathcal{H}_0] = 0$.  The lowest-order correction, $(1/2) [ S^{(1)} , \mathcal{H}_1]$, can be evaluated with the help of 

	\begin{equation}
	\begin{aligned}
	\frac{1}{1-\lambda \sigma^z} &= 
	\frac{1+\lambda \sigma^z}{1-\lambda^2},\quad (|\lambda| <1)\;,\\
	\frac{1}{1-\lambda (\sigma_i^z - \sigma_j^z)}
	&= 
	\frac{1-2\lambda^2}{1-4\lambda^2}
	+
	\frac{\lambda(\sigma_i^z - \sigma_j^z) - 2\lambda^2\sigma_i^z\sigma_j^z}{1-4\lambda^2}
	\;,
	\\
	&\quad\quad\quad\quad\quad\quad (|\lambda| <2)\;.
	\end{aligned}
	\end{equation}

	Using these identities and ignoring the small fluctuations $\delta B_j\sim \sigma_B$ in $\Delta_{ij}$, the leading perturbative correction is 
	\begin{widetext}
		\begin{equation}\label{APPeq:leadingcorr}
		\begin{aligned}
		\frac{1}{2}\left[S^{(1)},\mathcal{H}_1\right]
		&=
		-\frac{J\lambda}{8}\left(\sigma_1^z - \sigma_L^z\right)
		\\
		&\;\;
		-\frac{1}{16}\frac{J\lambda^2}{1-\lambda^2}
		\left(\kappa_{13}+\kappa_{L-2,L} + \sigma_1^z\left(\sigma_2^z - \sigma_3^z\right) - (\sigma_{L-2}^z - \sigma_{L-1}^z)\sigma_L^z - 2\sigma_1^z\sigma_3^z\kappa_{24} - 2\sigma_{L-2}^z \sigma_L^z\kappa_{L-3,L-1}\right)
		\\
		&\;\;
		+\frac{1}{16}\frac{J\lambda^2}{1-\lambda^2/4}
		\left((\sigma_1^z - \sigma_2^z)\sigma_3^z + \sigma_{L-2}^z( \sigma_{L-1}^z -\sigma_L^z) 
		-\kappa_{13} + \kappa_{L-2,L} - 2c_{12}c_{34} - 2c_{L-3,L-2}c_{L-1,L}
		\right)
		\\
		&\;\;
		-\frac{1}{16}\frac{J\lambda^2}{1-\lambda^2}
		\sum_{j=2}^{L-3}
		\left(\kappa_{j,j+2} + \sigma_j^z(\sigma_{j+1}^z - \sigma_{j+2}^z)- 2\sigma_j^z\sigma_{j+2}^z\kappa_{j+1,j+3}
		- 2c_{j-1,j}c_{j+1,j+2}\right)
		\\
		&\;\;
		-\frac{1}{16}\frac{J\lambda^2}{1-\lambda^2}
		\sum_{j=2}^{L-3}
		\left(\kappa_{j,j+2} - (\sigma_{j}^z - \sigma_{j+1}^z)\sigma_{j+2}^z- 2\sigma_j^z\sigma_{j+2}^z\kappa_{j-1,j+1}
		- 2c_{j,j+1}c_{j+2,j+3}\right)
		\\
		&\quad 
		+\mathcal{O}(J\lambda^3)
		\;,
		\quad\quad (\lambda \equiv J/g)
		\end{aligned}
		\end{equation}
	\end{widetext}
	where  $c_{ij} \equiv \sigma^+_i\sigma^-_j - \sigma^-_i\sigma^+_j$, 
	$\kappa_{ij} \equiv \sigma^+_i\sigma^-_j + \sigma^-_i\sigma^+_j$ are anti-Hermitian and Hermitian spin-flipping operators, and $\lambda \equiv J/g$.
	In the above expression, $\sigma_m^\alpha \equiv 0$ for $m\notin \{1,2,\cdots,L\}$ ($\alpha = \{z,\pm\}$). 
	The leading-order correction, given in the first line of Eq.~\eqref{APPeq:leadingcorr}, only involves the spins at the ends of the chain. One can also identify terms like $\sim J\lambda^2 c_{j,j+1}c_{j+2,j+3}$ that produce resonances \cite{Imbrie2016} when disorder is absent in $B_j$. For example, when $L=4$ one finds the terms $-\frac{J^3}{4g^2}\left(\sigma_{1}^-\sigma_{2}^+\sigma_{3}^+\sigma_{4}^-
	+
	\mathrm{H.c.}\right)$. This leads to the slow oscillation of $\Ket{\uparrow\downarrow\downarrow\uparrow} \leftrightarrow \Ket{\downarrow\uparrow\uparrow\downarrow}$ with a period that converges to $8\pi g^2/J^3$ in the limit of large $g/J$ in the undriven model without any field disorder ($\sigma_B = 0$), which ultimately limits the preservation of $\avg{\sigma_j^z}$.

	The Schrieffer-Wolff transformation of Eq.~\eqref{eq:SW} can be verified numerically by performing the transformation on the original Heisenberg model (with gradient field) and comparing the resulting spectrum with that of the corresponding Ising model.   This can be quantified by computing the average eigenvalue difference between the two Hamiltonians, $\eta = (1/2^N)\sqrt{\sum_i (E^{H}_i-E^{I}_i)^2}$.  Here $E^{H(I)}_i$ are the energy eigenvalues of the Heisenberg (Ising) model with gradient field.  Fig. \ref{fig:SWtest} shows that the two Hamiltonians approach each other at large gradients, with a $1/g$ dependence.

	\section{Perturbative Expansion for Quantum Fisher Information}\label{app:QFIpert}
	
		\begin{figure}
		\centering
		\includegraphics[scale = 0.65]{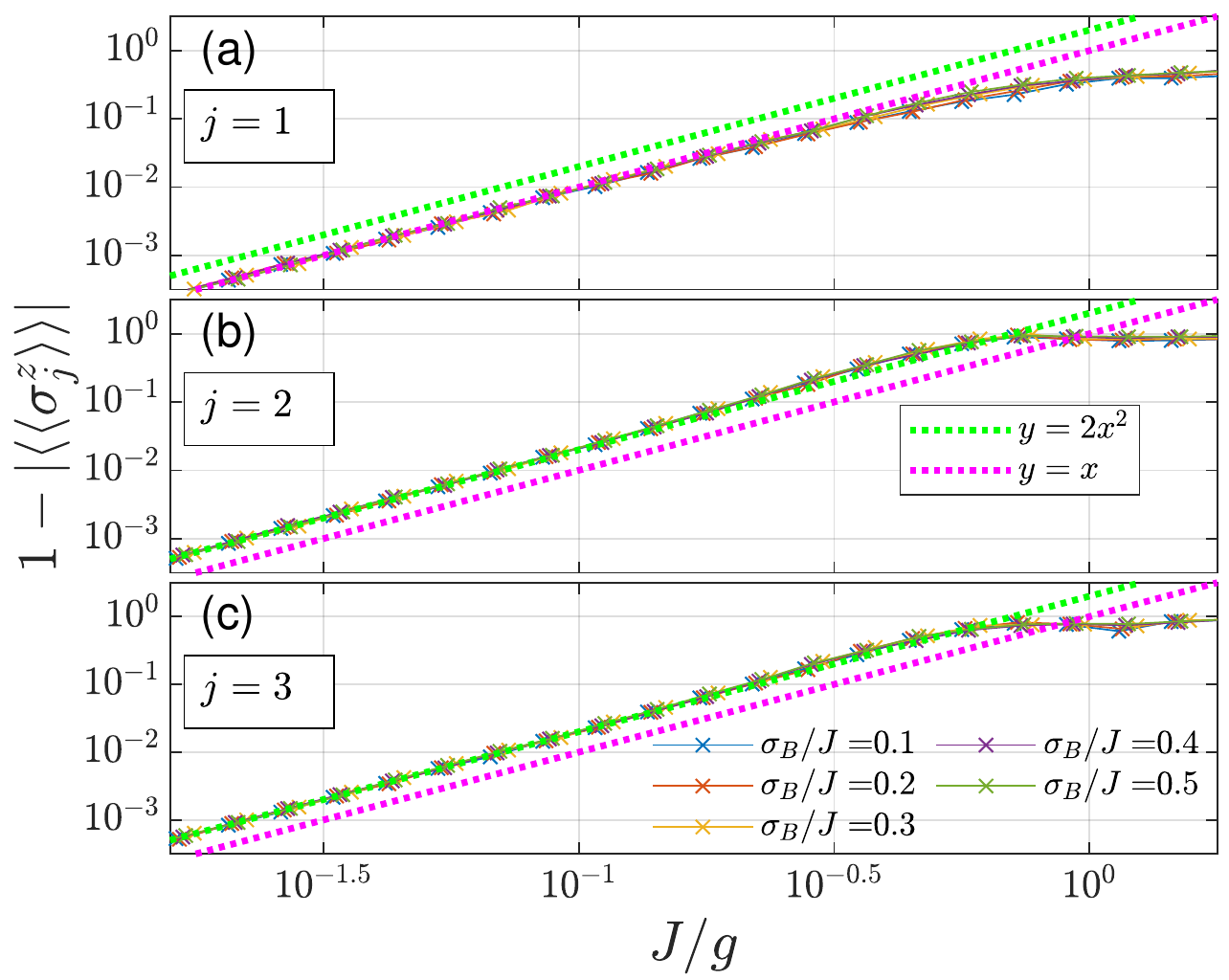}
		\caption{Numerical results (colored crosses) of one point functions and analytic predictions (dotted lines) for (a) an end spin and (b,c) two different bulk spins in an $L=6$ chain initialized in a N\'eel state. Individual data is averaged over the time span $10^3 J^{-1}<t<10^4 J^{-1}$ and $\sim 10^2$ disorder realizations. In all panels, we set $B_0/J=5\times 10^3$.}
		\label{fig:3loglines}
	\end{figure}
	
	\begin{figure}
		\centering
		\includegraphics[scale = 0.15]{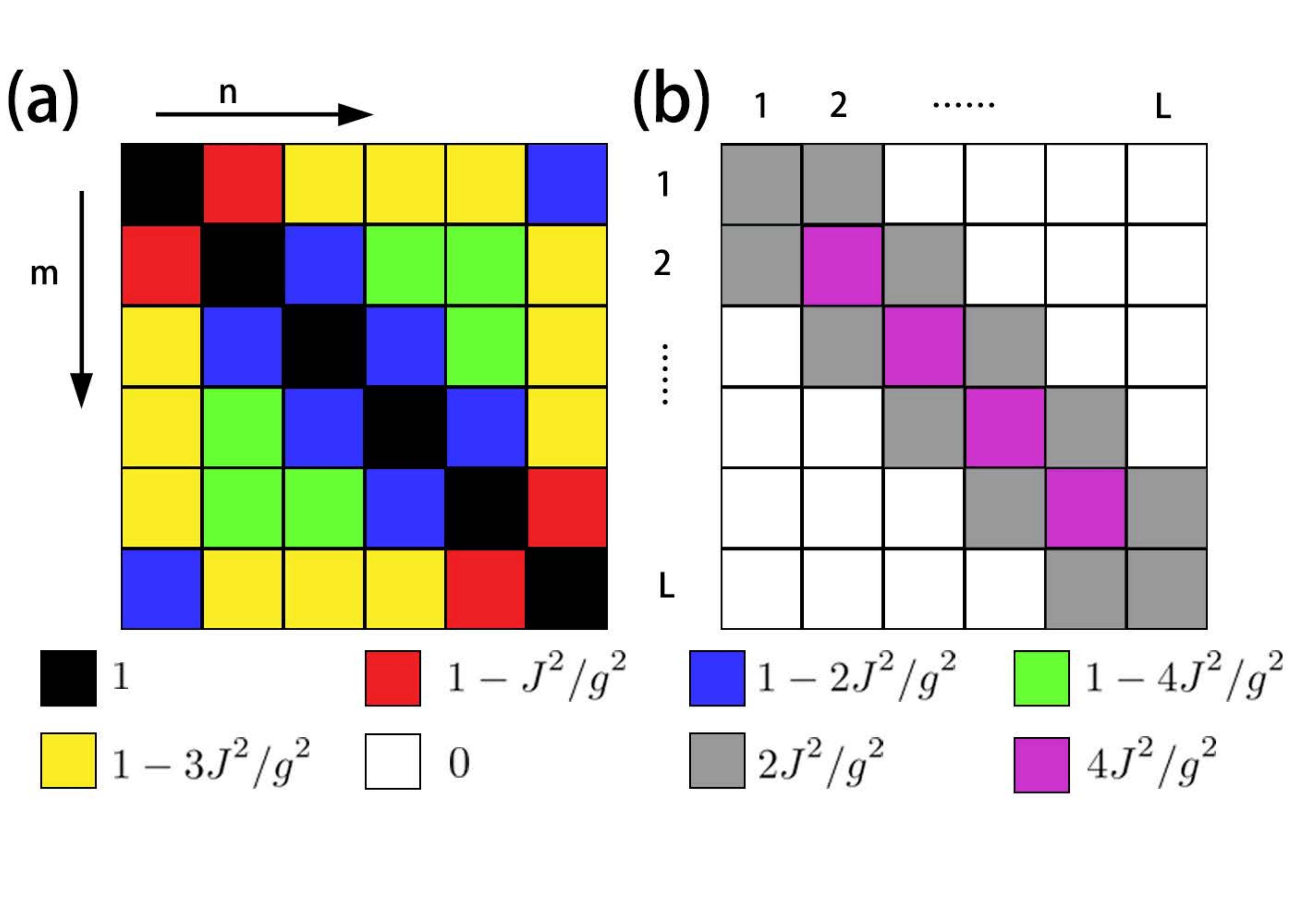}
		\caption{The leading-order terms of (a) the two point functions $(-1)^{m+n}{\avg{{\cal Z}_m{\cal Z}_n}}$ and (b) the correlators
			$(-1)^{m+n}(\avg{{\cal Z}_m{\cal Z}_n} - \avg{{\cal Z}_m}\avg{{\cal Z}_n})$.}
		\label{fig:MAT}
	\end{figure}
One may obtain an approximate analytic expression for the quantum Fisher information by computing the effect of the Schrieffer-Wolff transformation on local observables:
	\begin{equation}
	e^{S}\mathcal{O}_je^{-S} = \mathcal{O}_j + \mathrm{ad}_S\mathcal{O}_j + \frac{1}{2!}\mathrm{ad}_S^2\mathcal{O}_j +\cdots,
	\end{equation}
	where $\hbox{ad}_SX=[S,X]$. Consider $\mathcal{O}_j = \sigma_j^z$ and $e^S = e^{S^{(1)}}$. 
	From Eq. \eqref{APPeq:leadingcorr}, if $S^{(2)}$ exists, it has a lower order of $\lambda$: $S^{(2)}=\mathcal{O}(\lambda^3)$. 
	By ignoring $\mathcal{O}(\lambda^3)$ terms, we compute the truncated operator expansion:
	\begin{equation}\label{Appeq:Psi}
	{\cal Z}_j = \sigma_j^z + \mathrm{ad}_{S^{(1)}} \sigma_j^z + \frac{1}{2!}\mathrm{ad}^2_{S^{(1)}} \sigma_j^z,
	\end{equation}
	where
	\begin{equation}
	\begin{aligned}
	&\mathrm{ad}_{S^{(1)}}\sigma_j^z=
	2J
	\left[
	\frac{c_{j,j+1}}
	{2\Delta_{j,j+1} -J(\sigma_{j-1}^z-\sigma_{j+2}^z)}\right.
	\\
	&\qquad\qquad\left.
	-
	\frac{c_{j-1,j}}
	{2\Delta_{j-1,j} -J(\sigma_{j-2}^z-\sigma_{j+1}^z)}
	\right],
	\end{aligned}
	\end{equation}
	and
	\begin{equation}
	\begin{aligned}
	&\frac{1}{2!}\mathrm{ad}_{S^{(1)}}^2\sigma_j^z
	\\&=
	\frac{J^2}{2!}
	\left\{
	\frac{
		\sigma_j^{z}c_{j-1,j}}{\Delta_{j,j+1}^2}
	-\frac{1}{2}
	\left(\frac{\sigma_j^z -\sigma_{j+1}^z}{\Delta_{j,j+1}^2}
	+
	\frac{\sigma_j^z -\sigma_{j-1}^z}{\Delta_{j-1,j}^2}
	\right)
	\right.\\
	&
	\left.
	-\frac{1}{2}\left(
	\frac{\sigma_{j+1}^z c_{j,j+2}}
	{\Delta_{j,j+1}\Delta_{j+1,j+2}}
	+
	\frac{\sigma_{j-1}^z c_{j-2,j}}
	{\Delta_{j-2,j-1}\Delta_{j-1,j}}
	\right)
	\right\}
	+
	\mathcal{O}(\lambda^3).
	\end{aligned}
	\end{equation}
	One may then take the expectation for an initial N\'{e}el state $\Ket{\psi_i}=\Ket{\uparrow\downarrow\uparrow\downarrow\cdots}$, which yields
	\begin{equation}
	\begin{aligned}
	\avg{\mathrm{ad}_{S^{(1)}}\sigma_j^z}&=0,\\
	\avg{\{\sigma_j^z,\mathrm{ad}_{S^{(1)}}\sigma_j^z\}}&=0,\\
	\avg{\frac{1}{2!}\mathrm{ad}_{S^{(1)}}^2\sigma_j^z}&=\left\{
	\begin{aligned}
	&(-1)^{j+1}\lambda^2,\quad (j\ne 1,L)\\
	&(-1)^{j+1}\lambda^2/2,\quad (j= 1,L)
	\end{aligned}\right.,
	\\
	\avg{\{\sigma_j^z,\frac{1}{2!}\mathrm{ad}_{S^{(1)}}^2\sigma_j^z\}}
	&=2(-1)^{j}
	\avg{\frac{1}{2!}\mathrm{ad}_{S^{(1)}}^2\sigma_j^z},
	\end{aligned}
	\end{equation}
	where terms of order $\mathcal{O}(\lambda^3)$ are omitted.
	The above procedure requires a renormalization of the approximate state as follows:
	\begin{equation}\label{normalizeN}
	\begin{aligned}
	\mathcal{N}_j^2
	&= 
	\Bra{\psi_i}
	{\cal Z}_j^2
	\Ket{\psi_i}\\
	&=
	\Bra{\psi_i}
	1
	+ 
	\{\sigma_j^z,\mathrm{ad}_{S^{(1)}}\sigma_j^z\}
	+ 
	\{\sigma_j^z,\frac{1}{2!}\mathrm{ad}^2_{S^{(1)}}\sigma_j^z\}
	\\&\qquad\qquad
	+
	\left(\mathrm{ad}_{S^{(1)}}\sigma_j^z\right)^2
	+\mathcal{O}(\lambda^3)
	\Ket{\psi_i}\\
	&\approx\left\{
	\begin{aligned}
	&1 - 2\lambda^2,\quad (j\ne 1,L)\\
	&1 - \lambda^2,\quad (j = 1,L)
	\end{aligned}\right.\;.
	\end{aligned}
	\end{equation}
	Consequently, the leading terms of the normalized expectation value of ${\cal Z}_j$ with respect to the N\'{e}el state are
	\begin{equation}
	\mathcal{N}^{-1}_j\avg{{\cal Z}_j}
	\approx
	\left\{
	\begin{aligned}
	&(-1)^{j+1}(1 - 2\lambda^2),\quad (j\ne 1,L)\\
	&(-1)^{j+1}(1 - \lambda^2),\quad (j = 1,L)
	\end{aligned}\right.
	\;,
	\end{equation}
	as is numerically verified in Fig.~\ref{fig:3loglines}.
	
		\begin{figure}
		\includegraphics[scale=0.5]{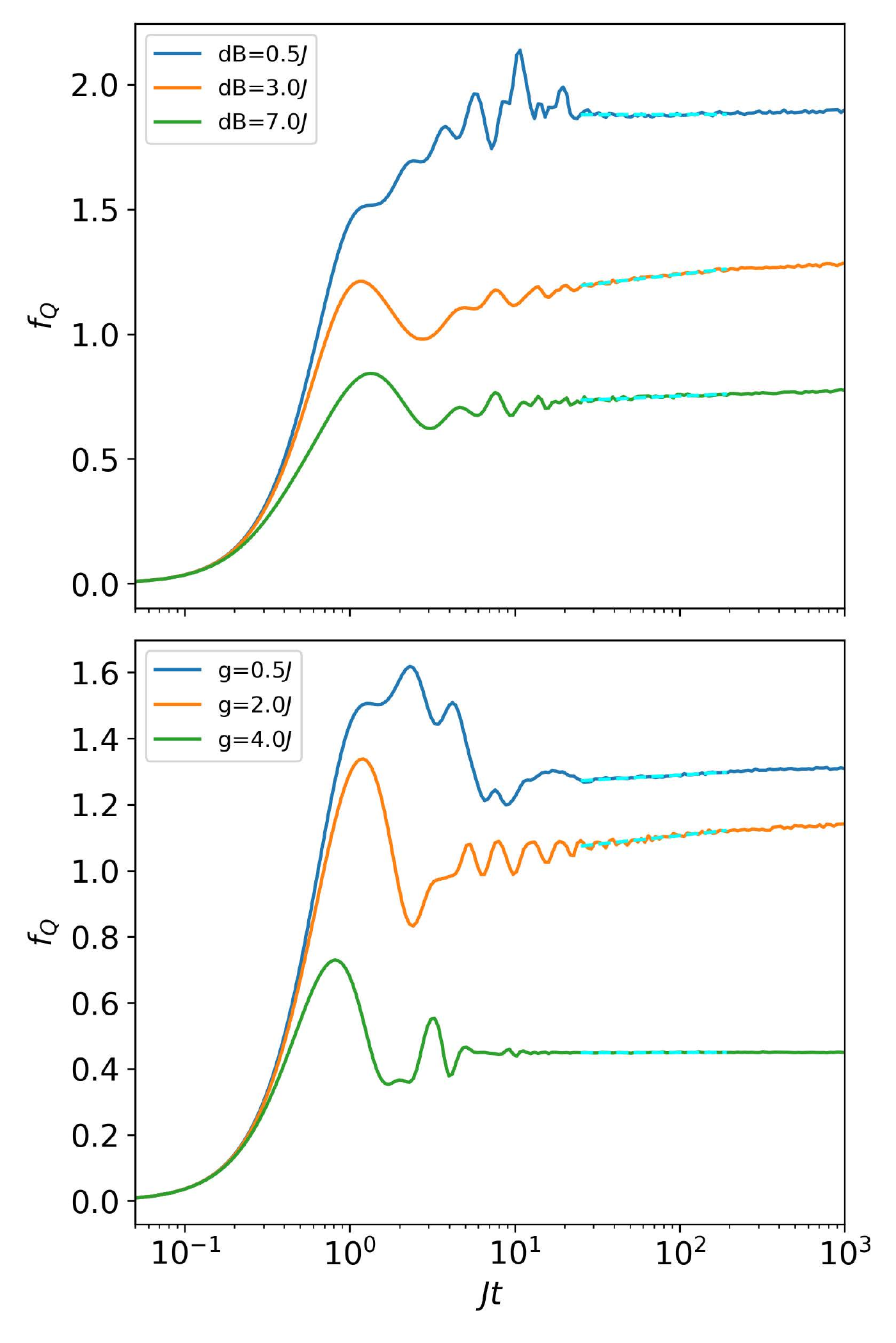}
		\caption{Quantum Fisher information as a function of time for (a) fixed field gradient $g=0$ and varying disorder strength $\delta B$; (b) fixed $\delta B = 0.5$ and varying field gradient.  Dashed blue lines indicate a logarithmic best fit to the QFI after the initial transient period and before saturation. Other parameters are $L=12$, $J=1$, $B_0=0$.  The initial state is the N\'{e}el state with $L=12$, and results are averaged over 4608 disorder realizations.  \label{fig:QFIvstappendix}}
	\end{figure}
	
	Similarly, we expand and truncate ${\cal Z}_m{\cal Z}_n$ as
	\begin{equation}
	\begin{aligned}
	&\sigma_m^z\sigma_n^z
	+
	\sigma_m^z\mathrm{ad}_{S^{(1)}}\sigma_n^z
	+
	(\mathrm{ad}_{S^{(1)}}\sigma_m^z)\sigma_n^z
	\\
	&\qquad
	+
	(
	\mathrm{ad}_{S^{(1)}}\sigma_m^z)(\mathrm{ad}_{S^{(1)}}\sigma_n^z
	)
	+
	\frac{1}{2}\sigma_m^z \mathrm{ad}^2_{S^{(1)}}\sigma_n^z
	\\
	&\qquad
	+
	\frac{1}{2}(\mathrm{ad}^2_{S^{(1)}}\sigma_m^z)\sigma_n^z
	+
	\mathcal{O}(\lambda^3)
	\;.
	\end{aligned}
	\end{equation}
	One finds (omitting terms of $\mathcal{O}(\lambda^3)$)
	and for general $m\ne n$, that
	\begin{equation}
	\begin{aligned}
	\avg{\sigma_m^z\sigma_n^z} &= (-1)^{m+n},
	\\
	\avg{\sigma_m^z \mathrm{ad}_{S^{(1)}}\sigma_n^z}
	&=
	\avg{(\mathrm{ad}_{S^{(1)}}\sigma_m^z) \sigma_n^z}
	=0,
	\\
	\avg{(
		\mathrm{ad}_{S^{(1)}}\sigma_m^z)(\mathrm{ad}_{S^{(1)}}\sigma_n^z
		)}
	&=
	\left\{
	\begin{aligned}
	2\lambda^2,\quad &|m-n|=1\\
	0,\quad &|m-n|>1
	\end{aligned}
	\right.,
	\\
	\end{aligned}
	\end{equation}
	and for the remaining terms, we obtain the following:
	\begin{itemize}
		\item If $m, n\ne 1,\mathrm{or}\;L$, then
		\begin{equation}
		\begin{aligned}
		\avg{\frac{1}{2!}\sigma_m^z( \mathrm{ad}^2_{S^{(1)}}\sigma_n^z)}
		&
		=
		\avg{\frac{1}{2!}(\mathrm{ad}^2_{S^{(1)}}\sigma_m^z)\sigma_n^z}
		\\&=-(-1)^{m+n} \lambda^2\;.
		\end{aligned}
		\end{equation}
		\item If $m = 1,\mathrm{or}\;L$, then
		\begin{equation}
		\begin{aligned}
		\avg{\frac{1}{2!}\sigma_n^z( \mathrm{ad}^2_{S^{(1)}}\sigma_{m}^z)}
		&
		=
		\avg{\frac{1}{2!}(\mathrm{ad}^2_{S^{(1)}}\sigma_{m}^z)\sigma_n^z}
		\\&
		=-(-1)^{m+n} \lambda^2/2\;.
		\end{aligned}
		\end{equation}
	\end{itemize}
	For $m=n$,  $\mathcal{N}^{-2}_m{\cal Z}_m^2 = \mathbbm{1}$. Combining all of the results above, the normalized two-point function is given by:
	\begin{equation}
	\mathcal{N}^{-1}_m \mathcal{N}^{-1}_n \avg{{\cal Z}_m {\cal Z}_n}\;.
	\end{equation} 
	The results for these expectation values are summarized in Fig.~\ref{fig:MAT}.
	
	Finally, the mean value of the Quantum Fisher information for $f_Q(t\to \infty)$  is then given by
	\begin{equation}
	\begin{aligned}
	f_Q &= \frac{1}{L}\sum_{m,n=1}^L\mathcal{N}_m^{-1} \mathcal{N}_n^{-1}(-1)^{m+n}\left(
	\avg{{\cal Z}_m{\cal Z}_n} - \avg{{\cal Z}_m}\avg{{\cal Z}_n}
	\right)\\
	&=
	\left(8-8/L\right)\lambda^2 + \mathcal{O}(\lambda^4)\;.
	\end{aligned}
	\end{equation}
	By symmetry, only even powers of $\lambda$ appear.

	\section{MBL Signatures: Comparison with the Literature}\label{app:MBLcompare}

	\begin{figure}
		\includegraphics[scale=0.49]{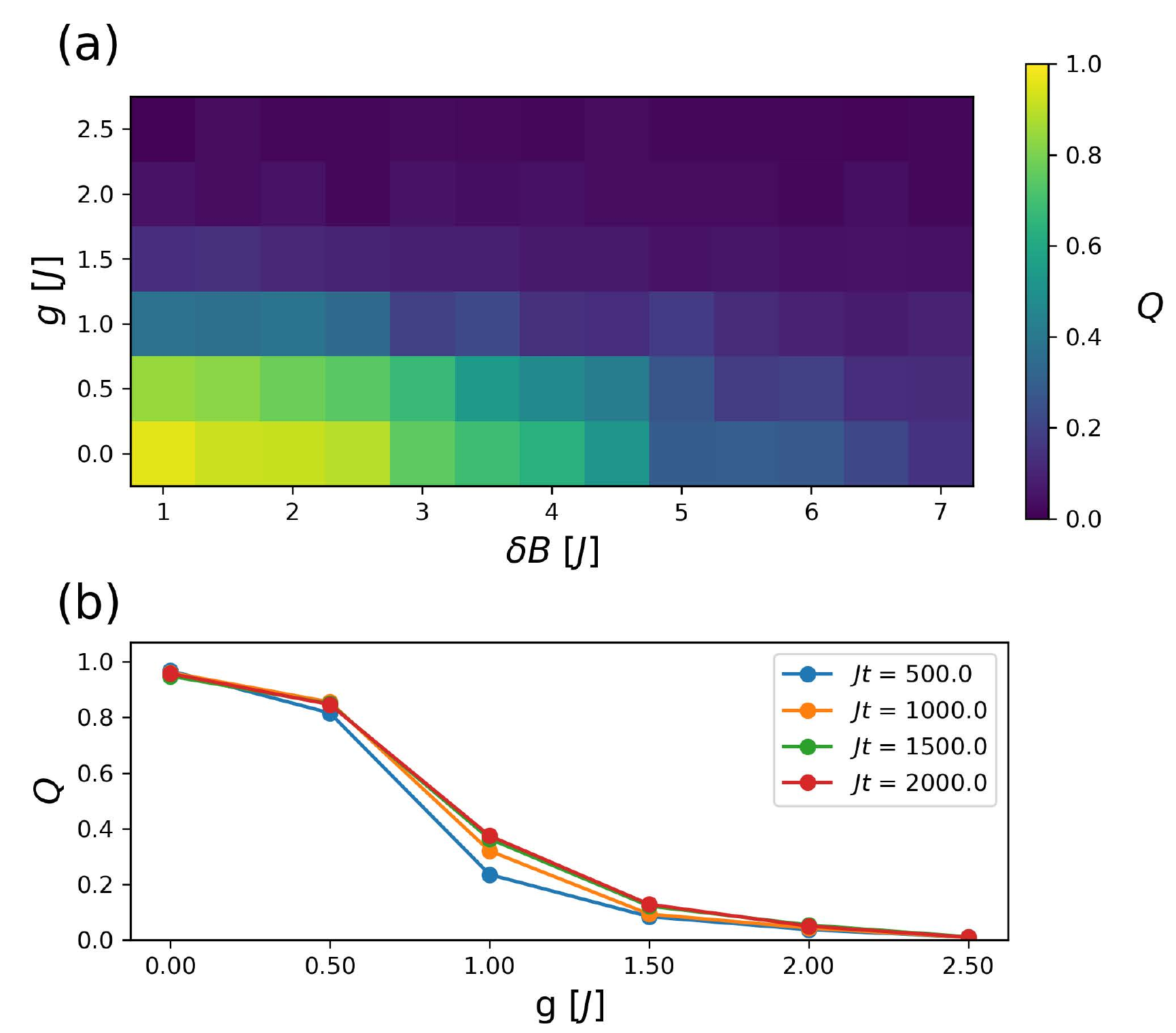}
		\caption{(a) Late-time dimensionless energy $Q$ as a function of gradient $g$ and disorder strength $\delta B$ under periodic driving.  For $Q \approx 1$ the system heats to infinite temperature, indicative of the ergodic phase.  For $Q \approx 0$ the system does not absorb energy, when the system is localized.  (b) $Q$ vs. $g$ for $\delta B = 1.0$, evaluated at different times. As $t$ increases, the curves approach fixed values $Q<1$, in constrast to the expectation for prethermalization, where $Q$ would ultimately approach 1 at long times. The system parameters are $N=12$, $J\!=\!1$, $B_0\!=\!0$, with the initial state being the ground state.  The driving parameters are $A\!=\!J$, $T\!=\!1$, $\eta = 0.4$. For (a), the $Q$ values are obtained by averaging over the last 20 periods for simulations of length $2000T$.  Results are averaged over 30 disorder realizations. \label{fig:heatingappendix}}
	\end{figure}
	
	Here we present further results for the MBL phase in the gradient field model, with the goal of comparing more directly with previous work \cite{Schulz2019,vanNieuwenburg2019}.  To this end, in the present section we sample the local field disorder from the uniform distribution $[-\delta B, \delta B]$, as in Ref. \cite{vanNieuwenburg2019}.
	
	In Fig. \ref{fig:QFIvstappendix} we present the normalized QFI as a function of time for different values of the disorder strength $\delta B$ [Fig. \ref{fig:QFIvstappendix}(a)] and field gradient [Fig. \ref{fig:QFIvstappendix}(b)].  In the former case, we take the field gradient $g=0$, while the latter has fixed disorder $\delta B = 0.5$.  As in Fig.~\ref{fig:QFIvst}, after an initial transient regime, the QFI stays relatively constant out to late times.  For small $\delta B$, the system is in the ergodic phase, characterized by a large late-time QFI, indicating the presence of entanglement.  As $\delta B$ or $g$ increases, the overall magnitude of the QFI is suppressed, suggesting a decrease in the entanglement as one enters the many-body localized phase.  We note that unlike Ref. \cite{Smith2016}, the QFI for our model does not display strong logarithmic growth, but instead is weakly logarithmic only near the transition from the ergodic to the MBL phase (orange curves in Fig. \ref{fig:QFIvstappendix}).
	
	As seen in Fig.~\ref{fig:heatingappendix}(a), the system only heats up ($Q \approx 1$) in the ergodic phase where $\delta B$ and $g$ are both small, with a transition line that is broadly consistent with the level statistics and participation ratio diagnostics \cite{vanNieuwenburg2019}.  In contrast, the MBL region of the phase diagram shows very little heating ($Q \approx 0$).  One may ask whether the failure of the system to heat up extends out to arbitrarily long times, or is rather due to the existence of a prethermal phase.  In the latter case, one may expect that curves of $Q(g)$, plotted for successively later times, would ultimately approach $Q(g) \approx 1$.  Fig.~ \ref{fig:heatingappendix}(b) reveals that this is not case: the $Q$ results approach a fixed curve, with $Q(g) \neq 1$, thus confirming that the gradient-induced localization is in fact MBL, as opposed to a prethermal phase.


%

\end{document}